\RequirePackage{fix-cm}
\documentclass[twocolumn]{svjour3}          %

\smartqed  %
\usepackage{xcolor}
\usepackage{graphicx}
\usepackage{times}
\usepackage{mathptmx}      %
\usepackage{textcomp}
\DeclareFontFamily{OT1}{pzc}{}
\DeclareFontShape{OT1}{pzc}{m}{it}{<-> s * [1.10] pzcmi7t}{}
\DeclareMathAlphabet{\mathpzc}{OT1}{pzc}{m}{it}

\usepackage{amssymb,amsfonts,amsmath,amscd}
\usepackage{bm} 

\usepackage{lipsum}

\usepackage[all]{nowidow}

\PassOptionsToPackage{hyphens}{url}
\usepackage{url,hyperref,lineno,microtype,subcaption}

\newcommand{\orcid}[1]{\href{https://orcid.org/#1}{\includegraphics[width=2ex,height=2ex]{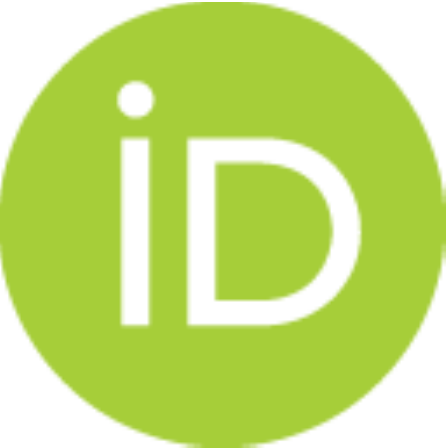}  orcid.org/#1}}

\usepackage{booktabs}%
\usepackage{tabu}%
\usepackage{longtable}%
\usepackage{rotating,tabularx}
\usepackage{multirow}
\usepackage{threeparttable}

\usepackage{siunitx}
\makeatletter
\let\oldtheequation\theequation
\renewcommand\tagform@[1]{\maketag@@@{\ignorespaces#1\unskip\@@italiccorr}}
\renewcommand\theequation{(\oldtheequation)}
\makeatother

\newcommand{\tcomma}{\text{, }}
\newcommand{\tdot}{\text{.}}
\newcommand{\tand}{\text{ and }}

\newcommand{\twith}{\text{ with }}

\newcommand{\vect}[1]{\bm{#1}}%
\newcommand{\mat}[1]{\bm{#1}}%
\newcommand{\bbm}{\begin{bmatrix}}
	\newcommand{\ebm}{\end{bmatrix}}

\newcommand{\T}{^\top}%

\newcommand{\fun}[1]{\mathrm{#1}}
\newcommand{\funarg}[2]{\fun{#1}\!\left(#2\right)}

\newcommand{\Proj}{\fun{\Pi}}

\newcommand{\set}[1]{\left\lbrace#1\right\rbrace}
\newcommand{\range}[1]{\left[#1\right]}
\newcommand{\norm}[1]{\left\Vert#1\right\Vert}
\newcommand{\abs}[1]{\left\vert#1\right\vert}

\newcommand{\convsym}{\text{\,\textasteriskcentered~}}
\newcommand{\convp}[2]{#1\convsym#2}

\newcommand{\nump}[2]{\num[round-mode=places,round-precision=#1]{#2}}
\newcommand{\focal}{f}
\newcommand{\Focal}{F}

\newcommand{\aperture}{A}%
\newcommand{\aperturevalue}{\mathrm{A\!V}}

\newcommand{\dof}{\mathrm{D\!O\!F}}

\newcommand{\focusdist}{h}
\newcommand{\Focusdist}{H}
\newcommand{\virtdepth}{\upsilon}%

\newcommand{\fnumber}{N}
\newcommand{\wfnumber}{N^{*}}

\newcommand{\blurpropcoef}{\kappa}

\newcommand{\internalparam}{\Omega}

\newcommand{\miradiusmm}{R}
\newcommand{\miradiuspix}{\tens{\varrho}}%
\newcommand{\blurradiusmm}{r}
\newcommand{\blurradiuspix}{\rho}

\newcommand{\distmlasensor}{\dist}
\newcommand{\distlensmla}{\Dist}

\newcommand{\distimg}{b}
\newcommand{\distobj}{a}

\newcommand{\lenscenter}{\vect{O}}

\newcommand{\mlinterdist}{\Delta\fmla}%
\newcommand{\mldiameter}{\mlinterdist}%
\newcommand{\mlcenter}{\vect{C}}
\newcommand{\mlindexes}{k,l}

\newcommand{\miinterdist}{\delta\fimg}%
\newcommand{\miinterdistmm}{\Delta\fimg}%
\newcommand{\micenter}{\vect{c}}
\newcommand{\miindexes}{\mlindexes}%

\newcommand{\mlinterdistapprox}{\lambda\miinterdistmm}

\newcommand{\principalpointx}{u_0}
\newcommand{\principalpointy}{v_0}

\newcommand{\mlpp}{\micenter{}^{\mlindexes}_0}
\newcommand{\mlppx}{\principalpointx^{\mlindexes}}
\newcommand{\mlppy}{\principalpointy^{\mlindexes}}

\newcommand{\pixelsize}{s}%

\newcommand{\Distor}{\fun{\varphi}}
\newcommand{\distor}[1]{\funarg{\varphi}{#1}}

\newcommand{\Psf}{\funarg{h}{x,y}}
\newcommand{\psf}{\fun{h}}

\newcommand{\Kthinlens}{\mat{K}\!}

\newcommand{\Transform}{\mat{T}}

\newcommand{\poseMLkl}{\Transform\fmla\!\left(\mlindexes\right)}
\newcommand{\poseLens}{\Transform\fcam}

\newcommand{\bapprojm}{\tens{P}\!} %

\newcommand{\fworld}{_w}
\newcommand{\fcam}{_c}
\newcommand{\fmla}{_\mu}

\newcommand{\fimg}{_i}

\newcommand{\fundistorted}{_u}

\newcommand{\mltype}[1]{{}^{\left(#1\right)}}
\newcommand{\mltypenb}{I}

\newcommand{\Img}{{I}}
\newcommand{\img}[1]{\Img\left(#1\right)}

\newcommand{\point}{\vect{p}}

\newcommand{\dist}{d}
\newcommand{\Dist}{D}

\newcommand*{\siecle}[1]{%
	\ifnum#1=1%
	\bsc{\romannumeral #1}\textsuperscript{er}~siècle%
	\else%
	\bsc{\romannumeral #1}\textsuperscript{e}~siècle%
	\fi
}
\newcommand{\resp}{\textit{resp.,~}}

\newcommand{\tfnumber}{$f$-number }

\newcommand{\feature}{\vect{p}}
\newcommand{\observation}{\feature{}^{n}_{\miindexes}}
\newcommand{\checkerboardcorner}{\point{}^{n}\fworld}

\usepackage[printonlyused]{acronym}

\acrodef{MFPC}{multi-focus plenoptic camera}
\acrodef{MLA}{micro-lenses array}
\acrodef{MIA}{micro-images array}
\acrodef{BAP}{Blur Aware Plenoptic}
\acrodef{MI}{micro-image}
\acrodef{MIC}{micro-image center}
\acrodef{SAI}{sub-aperture image}
\acrodef{CoC}{circle of confusion}
\acrodef{DoF}{depth of field}

\acrodef{DBSCAN}{density-based spatial clustering of applications with noise}
\acrodef{PnP}{Perspective-n-Point}
\acrodef{SfM}{Structure-from-Motion}
\acrodef{PIV}{particle image velocimetry}

\acrodef{RMSE}{root-mean-square error}
\acrodef{PSF}{point-spread function}

\newcommand{\ie}{i.e.,~}

\usepackage[
backend=bibtex, 
natbib=true, 
style=authoryear, 
doi=false, 
isbn=false, 
url=false, 
eprint=false, 
maxcitenames=1, 
mincitenames=1
]{biblatex}

\journalname{International Journal of Computer Vision}
\begin{document}
\title{
	Leveraging blur information for plenoptic camera calibration
\thanks{This work was supported by the AURA Region and the European Union (FEDER) through the MMII project of CPER 2015-2020 MMaSyF challenge.}
}

\author{Mathieu Labussi\`{e}re%
	\and C\'{e}line Teuli\`{e}re%
	\and Fr\'{e}d\'{e}ric Bernardin%
	\and Omar Ait-Aider%
}

\institute{%
	Mathieu Labussi\`{e}re$^1$ \at
	\orcid{0000-0001-8105-4139} \at
    \email{mathieu.labu@gmail.com}%
\and
	C\'{e}line Teuli\`{e}re$^1$ \at
	\orcid{0000-0002-8857-5524} \at
	\email{celine.teuliere@uca.fr}%
\and
	Fr\'{e}d\'{e}ric Bernardin$^2$ \at
	\orcid{0000-0002-1248-153X} \at
	\email{frederic.bernardin@cerema.fr}%
\and
	Omar Ait-Aider$^1$ \at
	\orcid{0000-0002-1711-187X} \at
	\email{omar.ait-aider@uca.fr}%
\and
$^1$~Universit\'{e} Clermont Auvergne, Clermont Auvergne INP, CNRS, Institut Pascal, F-63000 Clermont-Ferrand, France\\ 
$^2$~Cerema, \'{E}quipe-projet STI, 10 rue Bernard Palissy, F-63017 Clermont-Ferrand, France
}

\date{Received: date / Accepted: date}

\maketitle

\begin{abstract} %

This paper presents a novel calibration algorithm for plenoptic cameras, especially the multi-focus configuration, where several types of micro-lenses are used, using raw images only.
Current calibration methods %
rely on simplified projection models, use features from reconstructed images, or require separated calibrations for each type of micro-lens.
In the multi-focus configuration, %
the same part of a scene will demonstrate different amounts of blur according to the micro-lens focal length.
Usually, only micro-images with the smallest amount of blur are used.
In order to exploit all available data, we propose to explicitly model the defocus blur in a new camera model with the help of our newly introduced \acf{BAP} feature.
First, it is used in a pre-calibration step that retrieves initial camera parameters,
and second, to express a new cost function to be minimized in our single optimization process.
Third, it is exploited %
to calibrate the relative blur between micro-images.
It links the {geometric} blur, \ie the blur circle, to the {physical} blur, \ie the point spread function.
Finally, we use the resulting blur profile to characterize the camera's \acl{DoF}.
Quantitative evaluations in controlled environment on real-world data demonstrate the effectiveness of our calibrations.
\end{abstract}

\begin{figure}[b]
\centering
	\includegraphics[width=\linewidth]{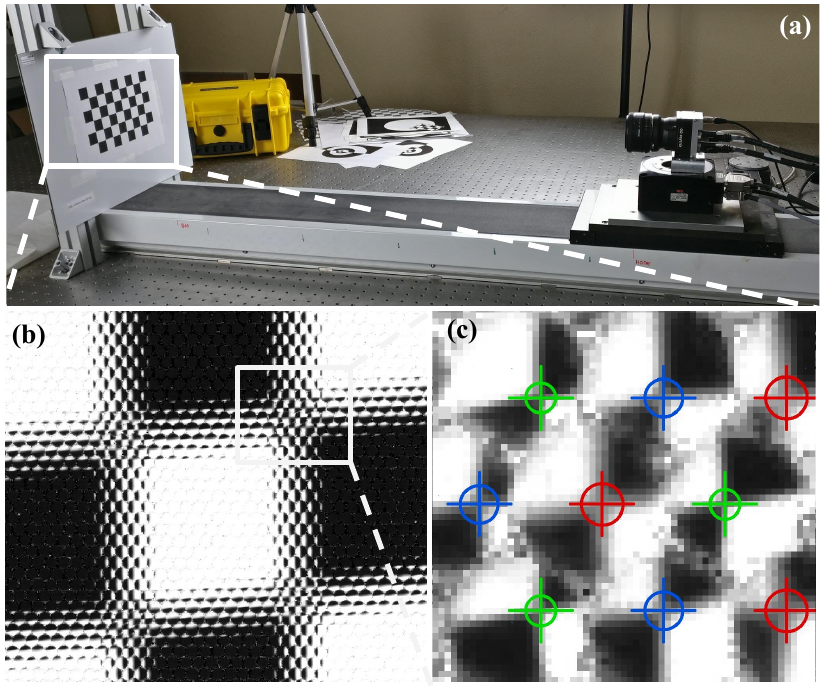}%
	\caption{
		The \texttt{Raytrix R12} multi-focus plenoptic camera used in our experimental setup \textbf{(a)}, along with a raw image of a checkerboard calibration target \textbf{(b)}. 
		The image is composed of several micro-images with different amounts of blur, arranged in a hexagonal grid.
		In each micro-image, our \acf{BAP} feature is illustrated by its center and its blur circle \textbf{(c)}.
	}%
	\label{fig:mfpc}\label{fig:setup}
\end{figure}

\acresetall

\section{Introduction}\label{sec:intro}

From \textit{Lumigraph} \citep{Lippmann1911b} to commercial \textit{plenoptic cameras} \citep{Ng2005b,Perwass2010c}, several designs have been proposed to capture information that cannot be captured by conventional cameras.
Said cameras capture only one point of view of a scene, whereas a \textit{plenoptic camera} is a device that allows to retrieve spatial as well as angular information.
A same point from a scene is projected into multiple observations on the sensor.
For instance, this redundant information can be used for digitally refocusing and rendering \citep{Bishop2012} or for depth estimation \citep{Johannsen2017b}.

This paper focuses on plenoptic cameras based on a \acf{MLA} placed between a main lens and a sensor as illustrated in \autoref{fig:model}.
The specific design of such a camera allows to multiplex both types of information onto the sensor in the form of a \acf{MIA}, as shown in \autoref{fig:mfpc}, but implies a trade-off between the angular and spatial resolutions \citep{Georgiev2006,Levin2008b,Georgiev2009e}.
It is balanced according to the \ac{MLA} position with respect to the main lens focal plane and the sensor plane, corresponding to \textit{unfocused} \citep{Ng2005b} or \textit{focused} \citep{Perwass2010c,Georgiev2012} configurations.

To further extend the \acf{DoF} of the plenoptic camera, a multi-focus configuration has been proposed by \citet{Perwass2010c,Georgiev2012}.
In this setup, the \ac{MLA} is composed of several micro-lenses with different focal lengths.
The same part of a scene will be more or less focused according to the micro-lens' type.
Usually, only micro-images with the smallest amount of blur are used.
Alternatively, specific patterns are used to exploit the information \citep{Palmieri2017}.
If one were able to relate the camera parameters to the amount of blur in the image, all information could be used simultaneously, without distinction between types of micro-lenses.
As a first step in that direction, we propose a calibration method that takes advantage of blur information.

Calibration is an initial step for applications using plenoptic imaging.
Conventional cameras are usually modeled as pinhole or thin lens. 
Due to the complexity of plenoptic cameras' design, the developed models are generally high dimensional.
Specific calibration methods have to be proposed to retrieve the intrinsic parameters of these models.

\subsection{Related work}

\paragraph{Unfocused plenoptic camera calibration.}
In the unfocused configuration, the main lens is focused at the \ac{MLA} plane and the sensor plane is placed at the \ac{MLA} focal plane. 
The \ac{MLA} %
is therefore focused at infinity, thus calling this configuration \textit{unfocused}.
The calibration of such plenoptic cameras \citep{Ng2005b} has been widely studied in the literature. %
Most approaches rely on a thin-lens model for the main lens and an array of pinholes for the micro-lenses.
\citet{Dansereau2013c} introduced a model to decode the pixels into rays, drawing inspiration from \citet{Grossberg2005}, for the \texttt{Lytro} plenoptic camera \citep{Ng2005b}.
Their model is not directly associated with physical parameters and is based on corner detection in reconstructed \acp{SAI}.
\citet{Zhou2019} proposed a practical two-step calibration method for unfocused plenoptic cameras.
Their model describes the camera physical parameters but still requires feature points extracted in reconstructed \acp{SAI}.
\citet{Bok2014} formulated a geometric projection model to estimate intrinsic and extrinsic parameters by utilizing raw images directly to avoid errors from reconstruction steps.
Their method includes analytical solution and non-linear optimization of the reprojection error of a novel line feature to overcome the difficulties in finding checkerboard corners. %
\citet{Shi2016} proposed a detailed model of a plenoptic camera in the context of \ac{PIV}.
Based on linear optics, they derived a model based on ray-tracing: contrarily to previous methods, 
they modeled the main lens and each micro-lens as thin-lenses.
\citet{Hahne2018b} developed a ray model by ray-tracing from the sensor side to the object space.
They consider only the chief ray, connecting \acp{MIC} to the exit pupil center.
\citet{OBrien2018} introduced a projection model used for their calibration method suited both for unfocused and focused plenoptic cameras.
They present a new feature called \textit{plenoptic disc}, similar in nature to the \ac{CoC} and defined by its center and its radius.
Their feature parametrization is in 3D and is in one-to-one correspondence with point positions in the camera frame, as it is detected in reconstructed image.
\citet{Zhao2020} recently presented a metric calibration method for unfocused plenoptic camera only also based on the \textit{plenoptic disc} but directly from raw image.

In summary, most of the above methods require reconstructed images (\acp{SAI}) to extract features,
and limit their model to the unfocused configuration, \ie setting the sensor plane at the micro-lens focal plane.
Therefore those models cannot be directly extended to the focused or multi-focus plenoptic camera.

\paragraph{Focused plenoptic camera calibration.}
With the arrival of commercial focused plenoptic cameras~\citep{Lumsdaine2009b,Perwass2010c}, new calibration methods have been proposed.
In this configuration, the micro-lenses focus on an intermediate image plane.
\citet{Johannsen2013b} formulated a general reprojection model in terms of the physical parameters of a \texttt{Raytrix} camera~\citep{Perwass2010c}.
They proposed a metric calibration and distortions correction using a grid of circular patterns.
This work considered a relatively simple model of lens distortion and required careful initialization of the optimization to converge due to high sensibility to local minima. 
\citet{Heinze2016b} improved the previous model by considering more sophisticated models of the main lens distortions.
They introduced new parameters including the tilt and shift for the main lens. 
They are able to distinguish each micro-lens type, calibrating then the distance between the \ac{MLA} and the sensor for each one but in separated calibration processes.
The projection model and the metric calibration procedure are incorporated in the \texttt{RxLive} software of \texttt{Raytrix GmbH}.
\citet{Strobl2016} presented a step-wise calibration approach to overcome the fragility of the initialization which hinders the final optimization. 
They first determined main lens parameters, then estimated \ac{MLA} parameters.
However, their calibration framework relied on reconstructed total focus images.
\citet{Zeller2014} introduced two new methods to calibrate a focused plenoptic camera and depth images obtained from it.
In further works \citep{Zeller2016a,Zeller2016h}, they improved the camera projection model by modeling the main lens as a thin lens instead of a pinhole.
The calibration process uses the reconstructed total focus image and virtual depth map to compute 3D observations.
All previous methods rely on reconstructed images (\acp{SAI}), which can lead to the introduction of errors in the reconstruction step as well as in the calibration process.
Usually, computation of reconstructed images requires camera parameters and/or depth information to avoid artifacts and reconstruction error.
To overcome this chicken and egg problem, several calibration methods focus on using only raw plenoptic images. %
\citet{Zhang2016a} proposed a calibration method based directly on observations from raw images.
They used a parallel bi-planar checkerboard to have a depth-scale prior. 
They considered a detailed model of the \ac{MLA} geometry that accounts for non-planarity of the array.
\citet{Zhang2018a} presented a multi-projection-center model based on the two planes parametrization~\citep{Levoy1996}.
They derived a calibration algorithm based on this model and projective transformation, suitable for both unfocused and focused plenoptic cameras. 
\citet{Noury2017b} presented a more complete geometrical model than the previous works.
This model relates 3D points to their corresponding image projections, working directly with raw images. 
They developed a new detector to find checkerboard corners with sub-pixel accuracy in each micro-image. %
They introduced a new cost function based on reprojection errors of both checkerboard corners and micro-lens centers in raw image space.
This enforces projected micro-lens centers to get closer to their corresponding \acp{MIC},
and makes their method robust to wrong parameters initialization especially concerning those of the \ac{MLA}.
However, their method does not consider different types of micro-lenses and forces them to act as pinholes. 

Several methods can account for the multi-focus setting.
\citet{Bok2017} extended their previous model \citep{Bok2014} to work with the focused plenoptic camera. 
They did not explicitly model the micro-lens focal lengths but introduced two additional intrinsic parameters that account for the \ac{MLA} setting.
Each setting -- one for each type of micro-lenses --, models a different distance between the \ac{MLA} and the sensor.
Their method can retrieve different intrinsics by running the optimization for each type separately.
\citet{Nousias2017} considered the geometric calibration of multi-focus plenoptic cameras.
Their method allows to identify the micro-lens types and their spatial arrangement. %
It operates on checkerboard corners retrieved by a custom micro-image corner detector.
Then, they applied their method on each type of micro-lens independently to retrieve specific intrinsic and extrinsic parameters for each configuration.
Latter researches \citep{Bok2017,Nousias2017,Noury2017b} have achieved improved performance through automation and accurate identification of feature correspondences in raw images. 
More recently, \citet{Wang2018} proposed a geometric calibration method for focused plenoptic cameras based on virtual image points, establishing the mapping from object points behind the main lens and the \ac{MLA} to image points on the sensor.
Their method can be extended to calibrate multi-focus cameras by considering each type of micro-lenses individually.

In conclusion, most of these methods rely on simplified models for optic elements: the \ac{MLA} misalignment is not considered, and the micro-lenses are modeled as pinholes thus not modeling their apertures.
Some do not consider distortions of the main lens or restrict themselves to the focused case.
Finally, few have considered the multi-focus case \citep{Heinze2016b,Bok2017,Nousias2017,Wang2018} but dealt with it in separate processes, leading to intrinsic and extrinsic parameters that vary depending on the type of micro-lens. 
\subsection{Contributions}

We present a new calibration method for plenoptic cameras.
To the best of our knowledge, it is the first to allow to calibrate the multi-focus plenoptic camera within a single process taking into account all types of micro-lenses simultaneously.
To exploit all available information, we propose to explicitly include the defocus blur in a new camera model.
Thus, we introduce a new \acf{BAP} feature defined in raw image space that enables us to handle the multi-focus case.
We present a new pre-calibration step using \ac{BAP} features from white images to provide a robust initial estimation of camera parameters.
We use our \ac{BAP} features in a single optimization process that retrieves intrinsic and extrinsic parameters of a \acl{MFPC} directly from raw plenoptic images of a checkerboard target.

\begin{figure*}[!h]
	\centering
	\includegraphics[width=\linewidth]{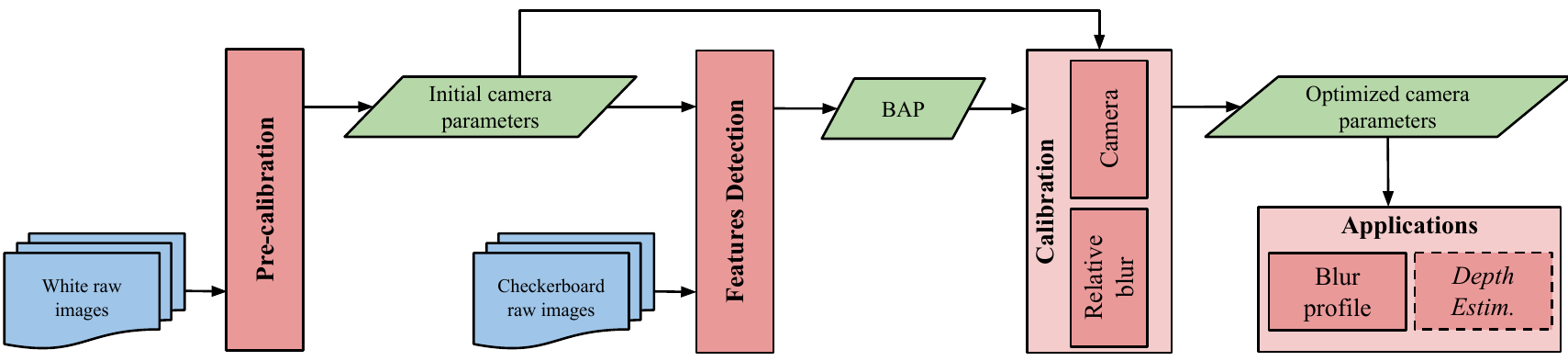}%
	\caption{
		Overview of our proposed method: first, the pre-calibration step retrieves initial camera parameters from white raw images at different apertures; then followed by the detection of \ac{BAP} features that are used by the camera calibration process and calibration ot the relative blur; %
		finally, once the camera is calibrated, it can be used, as addressed here, for profiling the camera, \ie to characterize the working range of the camera.
		Other applications can be considered, such as metric depth estimation.
	}%
	\label{fig:overview}
\end{figure*}

This paper extends our previous work \citep{Labussiere2020blur}.
In addition to our former contributions, we present here an ablation study of the camera parameters and add further comparisons with state-of-the-art calibration methods. 
{A new camera setup has also been tested to validate the generalization of our method, and a simulation setup is proposed to evaluate our method on \texttt{Lytro}-like configuration.} 
Moreover, we take advantage of our \ac{BAP} features to develop a new relative blur calibration process to link the geometric blur to the physical blur, \ie the \acf{CoC} to the \acf{PSF}.
This enables us to fully take advantage of blur in image space.
Finally, we propose to use the blur to profile the plenoptic camera in terms of \acf{DoF}.

\subsection{Paper organization}

\begin{figure*}[!t]
\centering
	\includegraphics[height=5.2cm]{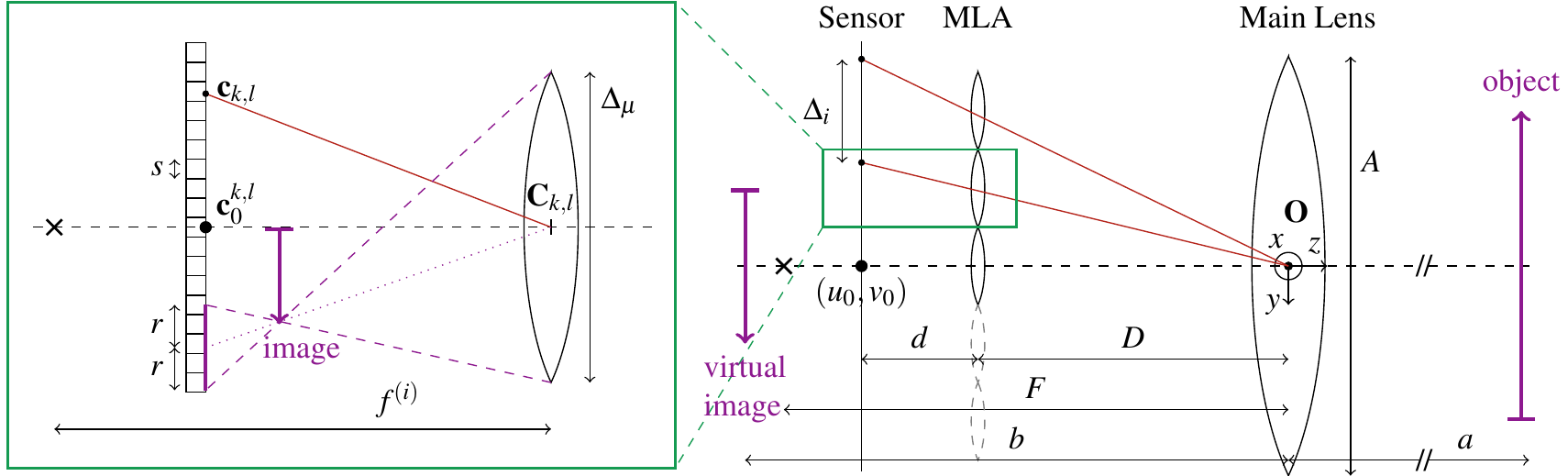}%
	\caption{
		Focused plenoptic camera model in {Galilean} configuration with the notations used in this paper. Object points are projected by the main lens behind the \ac{MLA} into a virtual intermediate space, and then re-imaged by each micro-lens onto the sensor.%
	}%
	\label{fig:notations}\label{fig:model}%
\end{figure*}

An overview of our method is given in \autoref{fig:overview}.
The remainder of this paper is organized as follows.
First, we present the camera model and how we model blur with our \ac{BAP} feature in \autoref{sec:theory}.
Second, we explain in \autoref{sec:precalib} how we leverage raw white images in the proposed pre-calibration step to initialize camera parameters.
Then, we detail the feature detection in \autoref{sec:detection} and the calibration processes in \autoref{sec:calibration}, \ie the camera calibration and the relative blur calibration.
Our experimental setup is presented in \autoref{sec:setup}.
Finally, our results are given and discussed in \autoref{sec:results}.
The notations used in this paper are shown in \autoref{fig:notations}.
Pixel counterparts of metric values are denoted in lower-case Greek letters.
Bold font denotes vectors and matrices.

\section{Camera and blur models}\label{sec:theory}

\subsection{The (multi-focus) plenoptic camera}

	We consider the focused plenoptic camera, especially the multi-focus case as described by \citet{Georgiev2012,Perwass2010c}.
	The camera is composed of a main lens and photosensitive sensor with a \acf{MLA} in between, as illustrated in \autoref{fig:model}.
	The multi-focus configuration implies that the micro-lenses array consists of $\mltypenb$ different types of lenses.
	The setup corresponds to the multi-focus system described by \citet{Perwass2010c} with $\mltypenb=3$.
	Note that our model can be applied to the single-focus plenoptic camera as well, corresponding then to the case where $\mltypenb = 1$.
	Finally, %
	the unfocused configuration is a special case of our model where
	the micro-lens focal length is equal to the distance between the \ac{MLA} and the sensor, \ie $\focal = \dist$.
					
	\subsubsection{Main Lens}
	
	The main lens is modeled as a thin-lens and maps an object point to a virtual point in an intermediate space called the virtual space. 
	An object at distance $\distobj$ is then projected at a distance $\distimg$ given the focal length $\Focal$ according to the thin-lens equation
	\begin{equation}\label{eq:thinlens}
		\frac{1}{\Focal} = \frac{1}{\distobj} + \frac{1}{\distimg}\tdot
	\end{equation}
	The main lens principal point is expressed as $\bbm\principalpointx& \principalpointy\ebm\T$ in image space.
	We model the main lens as parallel to the sensor plane. 
	Deviations from this hypothesis will be compensated for by tangential distortion parameters.	
	Furthermore, we define our camera reference frame as the main lens frame, with $\lenscenter$ being the origin, the $z$-axis coinciding with the optical axis and pointing outside the camera, and the $y$-axis pointing downwards.
	Distances are signed according to the following convention: $\Focal$ is positive when the lens is convergent; distances are positive when the point is real, and negative when virtual.
	
	\subsubsection{Distortions}
	
	We consider distortions of the main lens.
	Distortions represent deviations from the theoretical thin lens projection model.
	To correct those errors, we model the radial and tangential components of the lateral distortions using the model of Brown-Conrady \citep{Brown1966,Conrady1919}.
	Depth distortions have also been studied by \citet{Heinze2016b,Zeller2016a}, but \citet{Zeller2017,Noury2019} both empirically observed that the effects of depth distortions, for large focal length and for large object distance, can be neglected compared to stochastic noise of the depth estimation process.
	Therefore, we do not include depth distortion in our model.
	A distorted point $\point = \bbm x & y & z &1\ebm\T$ expressed in the main lens frame after projection (\ie in the virtual intermediate space) is thus transformed into $\point\fundistorted = \distor{\point} = \bbm x\fundistorted & y\fundistorted & z &1\ebm\T$ and is computed as
	\begin{equation}\label{eq:distortion}
	\left\{
	\begin{aligned}
	x\fundistorted =&~ x  \left(1 + Q_1 \varsigma^2 + Q_2 \varsigma^4 + Q_3 \varsigma^6 \right) &\text{[radial]}\\
	&+ P_1 \left(\varsigma^2 + 2 x{}^2\right) + 2 P_2 x y& \text{[tangential]}\\
	y\fundistorted =&~ y  \left(1 + Q_1 \varsigma^2 + Q_2 \varsigma^4 + Q_3 \varsigma^6\right) &\text{[radial]}\\
	&+ P_2 \left(\varsigma^2 + 2 y{}^2\right) + 2 P_1 x y& \text{[tangential]}\\
	\end{aligned}
	\right.
	\end{equation}
	where $\varsigma^2 = {x^2+y^2}$.
    The three coefficients for the radial component are given by 
    $\set{Q_1, Q_2, Q_3}$, 
	and the two coefficients for the tangential by 
	$\set{P_1, P_2}$.
	
	\subsubsection{Micro-lenses array}

	We also model the micro-lenses as thin-lenses allowing to take into account blur in the micro-image.
	The \ac{MLA} consists then of $\mltypenb$ different lens types with focal lengths $\focal\mltype{i}$ where $i \in [ 1\mathrel{{.}\,{.}}\nobreak \mltypenb]$ which are focused on $\mltypenb$ different planes.
	We make the hypothesis that all micro-lenses lie on the same plane.
	The \ac{MLA} is approximately centered around the optic axis.
	We define the farthest micro-lens along the $(-x)$-axis and the $(-y)$-axis as the origin of the \ac{MLA} frame, \ie the center of the upper-left micro-lens.
	The coordinates axes are orientated the same way as the ones of the main lens.
	The structural organization of the lenses can be an orthogonal or hexagonal %
	arrangement.
	The \ac{MLA} origin is at a distance $\Dist$ from the main lens and at a distance $\dist$ from the sensor.

	Furthermore, a detected \acf{MIC} usually does not coincide with the optical center of the considered micro-lens.		
	We take into account this deviation in opposition to orthographic projection of \acp{MIC} which causes inaccuracy in decoded light field. 
	Therefore, the principal point $\mlpp$ of the micro-lens indexed by $\left(\mlindexes\right)$ is given by
	\begin{equation}\label{eq:mlprincipalpoint}
	\mlpp = \bbm\mlppx\\\mlppy\ebm = \frac{\dist}{\Dist+\dist}\left(\bbm \principalpointx \\ \principalpointy \ebm - \micenter_{\mlindexes}\right) + \micenter_{\mlindexes}\tcomma
	\end{equation}
	where $\micenter_{\mlindexes}$ is the center of the micro-image $\left(\mlindexes\right)$ expressed in {pixel},
	as illustrated in \autoref{fig:notations}.

	\subsubsection{Micro-images array}
	
	Finally, each micro-lens produces a \acf{MI} onto the sensor. 
	The set of these micro-images has the same structural organization as the \ac{MLA}. 
	The data can therefore be interpreted as an array of micro-images, called by analogy the \acf{MIA}.
	The \ac{MIA} coordinates are expressed in image space.
	Let $\miinterdist$ be the pixel distance between two arbitrary consecutive micro-images centers $\micenter_{\mlindexes}$.
	With $\pixelsize$ the metric size of a pixel, let $\miinterdistmm = \pixelsize \miinterdist$ be its metric value,
	and $\mlinterdist$ be the metric distance between the two corresponding micro-lens centers $\mlcenter_{\mlindexes}$.
	From similar triangles, the ratio $\lambda$ between them is given by
	\begin{equation}\label{eq:mlinterdist}
	\lambda \triangleq \frac{\distlensmla}{\distmlasensor+\distlensmla} = \frac{\mlinterdist}{\miinterdistmm} \Longleftrightarrow \mlinterdist = \lambda \miinterdistmm = \frac{\distlensmla}{\distmlasensor+\distlensmla} \cdot \miinterdistmm\tdot
	\end{equation}
	We make the hypothesis that $\mlinterdist$ is equal to the micro-lens aperture.

	\subsubsection{Camera configuration}
	
	When the camera is in the \textit{unfocused} configuration, the distance separating the sensor and the \ac{MLA} is equal to the focal length of the micro-lenses, \ie $\distmlasensor = \focal$.
	Dealing with the focused plenoptic camera, we usually consider two possible configurations as presented by \citet{Georgiev2009d}: 1) Galilean, when objects are projected behind the image sensor; and 2) Keplerian, when objects are projected in front of the image sensor.
	When considering micro-lenses as thin-lenses, we have to take into account their focal lengths to configure the camera.
	In practice, considering an object projected at distance $\distimg$ by the main lens, four cases are possible but only two are able to produce an exploitable image, \ie with acceptable amount of blur, onto the sensor: $\distimg < \Dist$ and $\focal < \dist$ in Keplerian; and, $\distimg > \Dist$ and $\focal > \dist$ in Galilean.
	The condition $\distimg > \Dist$ can be achieved both when $\Focal > \Dist$ and $\Focal < \Dist$.
	The mode of operation is then constrained by the focal length of the micro-lenses, as suggested by \citet{Mignard-Debise2017}.
	We introduce then the definition of the \textit{internal} configuration according to the micro-lens focal length as 
	\begin{equation}\label{eq:configuration}
	\left\{
	\begin{aligned}
	&&\focal < \dist &\Longrightarrow& \text{Keplerian \textit{internal} configuration}\tcomma\\
	&&\focal > \dist &\Longrightarrow& \text{Galilean \textit{internal} configuration}\tdot
	\end{aligned}
	\right.
	\end{equation}

\subsection{Modeling blur within the plenoptic camera}

From optics geometry, the image of a point from a circular lens not focused on the sensor can be modeled by the \acf{CoC}.
Using a camera with a circular aperture, the blurred image is also circular in shape and is called the \textit{blur circle}. 
From similar triangles and from the thin-lens equation (\autoref{eq:thinlens}), the signed blur radius of the image of a point at a distance $\distobj$ from the lens is expressed as 
\begin{equation}\label{eq:blurradius}
\left\{
\begin{aligned}
&&\blurradiusmm &
=  \aperture\frac{\dist}{2}\left(\frac{1}{\focal} - \frac{1}{\distobj} - \frac{1}{\dist}\right) & \text{[metric]}\\
&&\blurradiuspix &= {\blurradiusmm}/{\pixelsize} &\text{[pixel]}\\
\end{aligned}
\right.		
\end{equation}
with 
$\pixelsize$ being the size of a pixel, and
$\aperture$ the {aperture} of this lens.
In continuous domain, the response of an imaging system to a not in-focus point, \ie the blur, can be expressed by the \acf{PSF}. 
Let $\img{x,y}$ be the observed blurred image of an object at a constant distance.
The image can be computed as the convolution of the \ac{PSF} noted $\Psf$, with the in-focus image, $\Img^*\!\left(x,y\right)$, such as
\begin{equation}\label{eq:psfconv}
\img{x,y} = \convp{\psf}{\Img^*\!\left(x,y\right)}\tcomma
\end{equation}
where $\convsym$ denotes the convolution operator.
If the lens {aperture} is circular and the level of blur low, the \ac{PSF} $\Psf$ can be efficiently modeled by a two-dimensional Gaussian given by
\begin{equation}\label{eq:psf}
\Psf = \frac{1}{2\pi\sigma^2}\exp\left(-\frac{x^2+y^2}{2\sigma^2}\right)\tcomma
\end{equation}
where the spread parameter $\sigma$ is proportional to the blur circle radius $\blurradiuspix$.
Therefore, we can write
\begin{equation}\label{eq:sigma2rho}
\sigma \propto \blurradiuspix \Leftrightarrow \sigma = \kappa \cdot \blurradiuspix
\end{equation}
where $\kappa$ is a camera constant that should be determined by calibration \citep{Pentland1987,Subbarao1989}.
Note that the spatially-variant spread parameter $\sigma$ thus depends on the object distance $\distobj$. %

The blur radius $\blurradiuspix$ appears at several levels within the camera projection:
in the blur introduced by the thin-lens model of the micro-lenses 
and in the formation of the micro-images while taking a white image.
Each micro-lens $\left(\mlindexes\right)$ projects virtual points onto the sensor at a position $\left(u,v\right)$, with a blur radius $\blurradiuspix$ depending on the distance to the point and the micro-lens type.

\subsection{BAP features and projection model}

To leverage this blur information, we introduce a new \acf{BAP} feature characterized by its center and its radius, noted $\feature = \bbm u& v& \blurradiuspix & 1\ebm\T$.
The BAP feature are visualized in \autoref{fig:mfpc}.
Therefore, our complete plenoptic camera model allows us to link a scene point $\point\fworld = \bbm x&y&z&1\ebm\T$ to our new \ac{BAP} feature $\feature$ in homogeneous coordinates through each micro-lens $\left(\mlindexes\right)$ such as
\begin{equation}\label{eq:completemodel}
\bbm u \\ v \\ \blurradiuspix \\ 1\ebm \propto
\bapprojm\left(i,\mlindexes\right)
\cdot 	\poseMLkl
\cdot 	\distor{
	\Kthinlens\left(\Focal\right)
	\cdot 	\poseLens
	\cdot 	\point\fworld
}\tcomma%
\end{equation}
where
$\bapprojm\left(i,\mlindexes\right)$ is the blur aware plenoptic projection matrix through the micro-lens $\left(\mlindexes\right)$ of type $i$, and computed as
\begin{align}\label{eq:bapprojm}
&\bapprojm\left(i,\mlindexes\right) = \mat{P}\left(\mlindexes\right)\cdot \Kthinlens\left(\focal\mltype{i}\right) \\
=& \bbm 
{\dist}/{\pixelsize} & 0 & \principalpointx^{\mlindexes} & 0 \\
0 & {\dist}/{\pixelsize} & \principalpointy^{\mlindexes} & 0 \\
0 & 0 & \frac{\mlinterdist}{2\pixelsize} & -\frac{\mlinterdist}{2\pixelsize}\dist \\
0 & 0 & -1 & 0		
\ebm 
\bbm 
1 & 0 & 0 & 0 \\
0 & 1 & 0 & 0 \\
0 & 0 & 1 & 0\\
0 & 0 & -1/\focal\mltype{i} & 1	
\ebm\tdot\nonumber
\end{align}
$\mat{P}\left(\mlindexes\right)$ is a matrix that projects the 3D virtual point onto the sensor and taking into account the blur radius.
$\Kthinlens\left(\focal\right)$ is the thin-lens projection matrix for the given {focal length}.
$\poseLens$ is the pose of the main lens with respect to the world frame 
and $\poseMLkl$ is the pose of the micro-lens $\left(\mlindexes\right)$ expressed in the camera frame.
The function $\funarg{\Distor}{\cdot}$ models the lateral distortions.

Finally, the projection model from \autoref{eq:completemodel} consists of a set $\Xi$ of $\left(16+\mltypenb\right)$	
intrinsic parameters to be optimized, including:
	the main lens {focal length} $\Focal$, expressed in $\Kthinlens\left(\Focal\right)$,
	and its five lateral distortion coefficients %
	$Q_1$, $Q_2$, $Q_3$, $P_1$, and $P_2$, expressed in $\distor{\cdot}$;
	the sensor translations, encoded in $\dist$ and $\left(\principalpointx, \principalpointy\right)$ through \autoref{eq:mlprincipalpoint}, from $\mat{P}\left(\mlindexes\right)$;
	the \ac{MLA} pose, including its three rotations $\left(\theta_x, \theta_y, \theta_z\right)$ and three translations $\left(t_x, t_y, \Dist\right)$, and 
	the micro-lens pitch $\mlinterdist$, expressed in $\poseMLkl$;
	and, the $\mltypenb$ micro-lens {focal lengths} $\focal\mltype{i}$, in $\Kthinlens(\focal\mltype{i})$.
\subsection{Profiling the depth of field of the plenoptic camera}

From calibrated camera parameters, we can compute the \acf{DoF} of each micro-lens type and the \textit{blur profile} -- the blur radii as function of the object distance --, in order to profile the plenoptic camera. 
The analysis can be done with respect to the \ac{MLA} pose, and then extended to object space by back-projection.
A point at a distance $\distobj$ from \ac{MLA} is projected back into object space at a distance $\distobj'$ according to the thin-lens equation through the main lens, such as
\begin{equation}\label{eq:mla2obj}
\distobj' = \frac{\left(\distlensmla - \distobj\right)\cdot\Focal}{\left(\distlensmla - \distobj\right) - \Focal}\tdot
\end{equation}
Let $r_0$ be the minimal acceptable radius of the \ac{CoC}.
The smallest diffraction-limited spot resolved by a lens in wave optics, \ie the radius of the first null of the Airy disc, is $r^*= 1.22 \cdot \nu \cdot \wfnumber$, where $\nu$ is the considered light wavelength, and $\wfnumber = \dist / \aperture$ is the working $f$-number of the lens.
The minimal acceptable radius is the maximum between this limit and half the size of a pixel, such as $r_0 = \max\left(r^*\tcomma {s}/2\right)$.
For a a micro-lens of type $(i)$, the focus plane distance is given by
\begin{equation}\label{eq:focusplane}
a_0^{(i)} = \left(\frac{1}{f^{(i)}} - \frac{1}{d} \right)^{-1} = \frac{df^{(i)}}{d-f^{(i)}}\tdot
\end{equation}
Let $\aperture$ be the micro-lens aperture, we derive then the \textit{far} $a_+$ and \textit{near} $a_-$ focus planes distances:
\begin{equation}\label{eq:farnearfocus}
\left\{
\begin{aligned}
	a_+^{(i)} &= \frac{dA \cdot a_0^{(i)}}{Af\mltype{i} - 2r_0\left(a_0^{(i)} - f\mltype{i}\right)} &&\text{~~~~[far]}\\
	a_-^{(i)} &= \frac{dA \cdot a_0^{(i)}}{Af\mltype{i} + 2r_0\left(a_0^{(i)} - f\mltype{i}\right)} &&\text{~~~~[near]}\tdot\\
\end{aligned}
\right.		
\end{equation}
The \ac{DoF} of a micro-lens of type $(i)$ is computed as the distance between the near and far focus planes, such as 
\begin{equation}\label{eq:dofml}
\dof^{(i)} = \left|{a_+^{(i)}}\right| - \left|{a_-^{(i)}}\right| = \frac{Af\mltype{i} \cdot a_0^{(i)} \cdot 2r_0\left(a_0^{(i)}-f\mltype{i}\right)}{\left(Af\mltype{i}\right)^2 - 4r_0^2\left(a_0^{(i)}-f\mltype{i}\right)^2}.
\end{equation}
Note that to fully exploit the combined extended \acp{DoF} without gaps, the micro-lenses \acp{DoF} should either just touch or slightly overlap \citep{Perwass2010c}.
Finally, under this consideration, the total \ac{DoF} of the plenoptic camera in \ac{MLA} space is computed using the micro-lenses \acp{DoF} as
\begin{equation}\label{eq:dofcamera}
\dof = \max_{i}\left\{ \left|a_+^{(i)}\right|\right\} - \min_{i}\left\{ \left|a_-^{(i)}\right|\right\}\tdot
\end{equation}
We can finally plot the \textit{blur profile} of the camera, %
along with the focal planes and the total \ac{DoF} %
as illustrated by the \autoref{fig:blurprofileobj}.

\section{Pre-calibration using raw white images}\label{sec:precalib}

The goal of the pre-calibration step is to provide a strong initial estimate of the camera parameters.
Inspired from \textit{depth from defocus} theory \citep{Subbarao1994}, 
we leverage blur information to estimate our blur radius by varying the main lens aperture and using the different micro-lenses focal lengths, in combination with parameters from the image space.
This is achieved by using raw white images acquired with a light diffuser mounted on the main objective, and taken at different apertures.
We then show how the blur radii are linked to camera parameters, thus enabling their initialization.
\subsection{Micro-images array calibration}

First, the \acf{MIA} is calibrated using raw white images.
We compute the micro-image centers $\set{\micenter_{\mlindexes}}$ by the intensity centroid method with sub-pixel accuracy \citep{Thomason2014,Noury2017b,Suliga2018}.
The distance between two micro-image centers $\miinterdist$ is then computed as the optimized edge-length of a fitted 2D regular grid mesh.
The optimization is conducted by non-linear minimization of the distances between the grid vertices and the corresponding detected \acp{MIC}.
The pixel translation offset in image coordinates, $(\tau_x,\tau_y)$,
and the rotation around the $\left(-z\right)$-axis, $\vartheta_z$,
are also determined during the optimization process.

\subsection{Deriving the micro-image radius}

\begin{figure}[t]
	\centering
	\includegraphics[width=\linewidth]{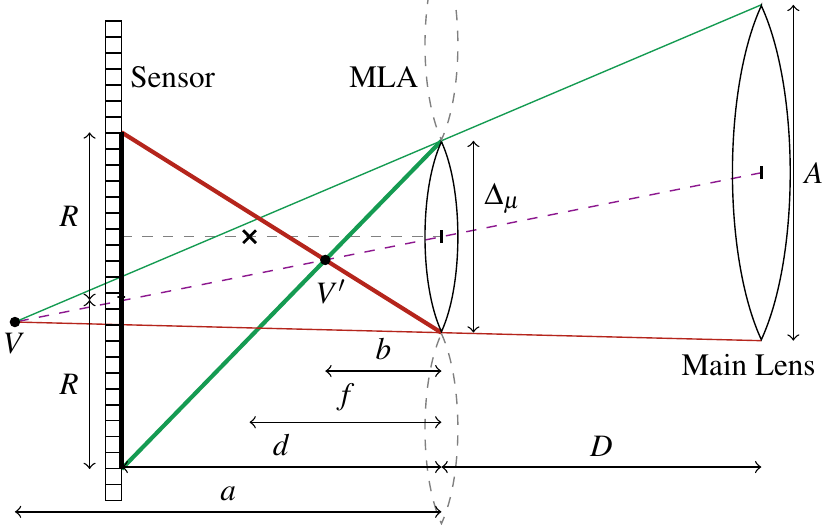}%
	\caption{%
		Formation of a micro-image with its radius $\miradiusmm$ through a micro-lens while taking a white image using a light diffuser, at an {aperture} $\aperture$, in Keplerian internal configuration.
		The point $V$ is the vertex of the cone passing by the main lens and the considered micro-lens.  
		$V'$ is the image of $V$ by the micro-lens and $\miradiusmm$ is the radius of its blur circle.	
	}%
	\label{fig:blurradius}%
\end{figure}

\begin{figure*}[t]
\centering
	\includegraphics[width=\linewidth]{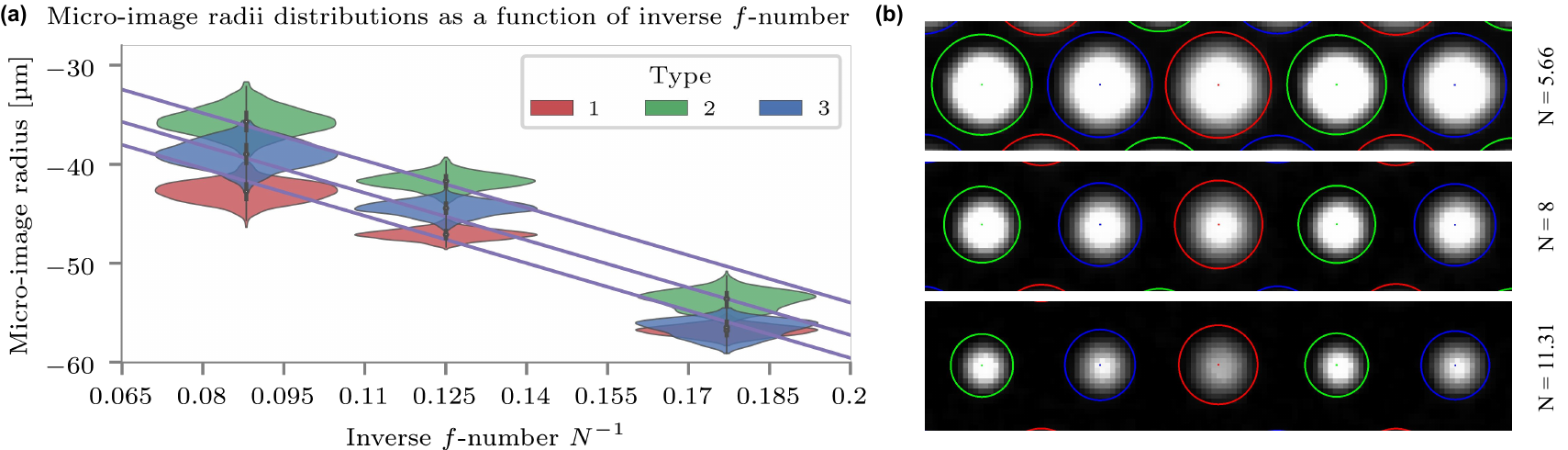}
	\caption{
		\textbf{(a)} Micro-image radii as function of the inverse $f$-number (in \textit{magenta}), with their distributions represented by the violin-boxes, for our camera consisting of $\mltypenb = 3$ different types.
		\textbf{(b)} Each type of micro-lens is identified by its color (type $(1)$ in \textit{red}, type $(2)$ in \textit{green}, and type $(3)$ in \textit{blue}) with its computed radius. 
	}\label{fig:radii}
\end{figure*}

In white images taken with a light diffuser and a controlled {aperture}, each type of micro-lens produces a \acf{MI} with a specific size and intensity.
This provides a mean to distinguish between them (\autoref{fig:radii}).
The process of capturing a white image is equivalent for the micro-lenses to imaging a white uniform object of diameter $\aperture$ at a distance $\Dist$.
The imaging process is schematized in \autoref{fig:blurradius}.
Using optics geometry, %
the image of this object, \ie the resulting \ac{MI}, corresponds to the image of an imaginary point $V$ constructed as the vertex of the cone passing through the main lens and the considered micro-lens.
Let $\distobj$ be the signed distance of this point from the \ac{MLA} plane, 
expressed from similar triangles and \autoref{eq:mlinterdist} as
\begin{equation}\label{eq:vdistmla}
\distobj = -\Dist\frac{\mldiameter}{\aperture-\mldiameter} 
= - {\Dist}\left(\aperture\left(\frac{\dist+\Dist}{\Dist}\cdot\frac{1}{\miinterdistmm}\right) - 1\right)^{-1}\tcomma
\end{equation}
with 
$\aperture$ being the main lens {aperture}.
Note the minus sign is added because the vertex is always formed behind the \ac{MLA} plane, and thus considered as a virtual object for the micro-lenses.
Geometrically, the \ac{MI} formed is the \textit{blur circle} of this imaginary point $V$.
Therefore, injecting the latter expression in \autoref{eq:blurradius}, the metric \ac{MI} radius $\miradiusmm$ is given by
\begin{align}\label{eq:miradiusaperture}
\miradiusmm &= \frac{\mldiameter}{2}\dist\left(\frac{1}{\focal} - \frac{1}{\distobj} - \frac{1}{d}\right) \nonumber\\ 
&= \left(\frac{\miinterdistmm\cdot\Dist}{\dist+\Dist}\right)\cdot\frac{\dist}{2}\cdot\left(\frac{1}{\focal} + \left(\aperture\left(\frac{\dist+\Dist}{\Dist}\cdot\frac{1}{\miinterdistmm}\right) - 1\right) \frac{1}{\Dist} - \frac{1}{d}\right) \nonumber \\
&= \aperture\cdot \frac{\dist}{2\Dist} +  \left(\frac{\miinterdistmm\cdot\Dist}{\dist+\Dist}\right) \cdot \frac{\dist}{2} \cdot \left(\frac{1}{\focal} - \frac{1}{\Dist} - \frac{1}{d}\right)
\tdot
\end{align}
From the above equation, the \ac{MI} radius $\miradiusmm$ depends linearly on the {aperture} of the main lens.
However, the main lens {aperture} cannot be measured directly whereas we have access to the \tfnumber value.
Recall that the \tfnumber of an optical system is the ratio of the system's {focal length} $\Focal$ to the aperture, $\aperture$, given by
$\fnumber = {\Focal}/{\aperture}$.
Finally, we can express the \ac{MI} radius for each micro-lens {focal length} type $i$ as
\begin{equation}\label{eq:miradiusfnumber}
\funarg{\miradiusmm_i}{\fnumber^{-1}} = m \cdot \fnumber^{-1} + q_i
\end{equation}
with
\begin{equation}
m = \frac{\dist\Focal}{2\Dist} \text{~~~and~~~}
q_i = \frac{1}{\focal\mltype{i}} \cdot \left(\frac{\miinterdistmm\cdot\Dist}{\dist+\Dist}\right) \cdot \frac{\dist}{2}  - \frac{\miinterdistmm}{2} \tdot\label{eq:miradiusfnumbermc}
\end{equation}%
We thus relate the \ac{MI} radius to the plenoptic camera parameters.
It is a function of fixed parameters ($\dist, \Dist, \Focal$), measured parameters ($\miinterdistmm = \pixelsize\cdot\miinterdist$) and variable parameters ($\fnumber$ and $\focal\mltype{i}$ with $i \in [ 1\mathrel{{.}\,{.}}\nobreak \mltypenb]$).

\noindent Let $\internalparam$ be the set of parameters $\set{m, q'_1, \dots, q'_\mltypenb}$, where  $q_i'$ is the value obtained by 
\begin{equation}\label{eq:qprime}
q_i' = \frac{1}{\focal\mltype{i}} \cdot \left(\frac{\miinterdistmm\cdot\Dist}{\dist+\Dist}\right) \cdot \frac{\dist}{2} = q_i + \frac{\miinterdistmm}{2}\tdot
\end{equation}
They are used to compute the radius part of the \ac{BAP} feature and to initialize the camera parameters.

\paragraph{Micro-image radii estimation.}
From raw white images, we measure each \ac{MI} radius $\miradiuspix = \abs{\miradiusmm}/\pixelsize$ in \si{pixel} based on image moments fitting.
We use the second order central moments of the micro-image to construct a covariance matrix.
The radius $\miradiuspix$ is proportional to the computed standard deviation $\sigma$.
Recall that raw moments and centroid are given by 
\begin{align*}
M_{{ij}}=\sum _{x,y}x^{i}y^{j}\img{x,y} &&\tand&&  {\displaystyle \{{\bar {x}},\ {\bar {y}}\}=\left\{{\frac {M_{10}}{M_{00}}},{\frac {M_{01}}{M_{00}}}\right\}}\tcomma
\end{align*}
and the central moments by
\begin{equation}
\mu _{{pq}}=\sum _{{x, y}}(x-{\bar  {x}})^{p}(y-{\bar  {y}})^{q}\img{x,y}\tdot
\end{equation}
The covariance matrix is then computed as
\begin{equation}
\mathrm{cov}\left[\img{x,y}\right] = \frac{1}{\mu_{{00}}} \bbm \mu_{{20}} & \mu_{{11}} \\ \mu_{{11}} & \mu_{{02}}\ebm = \bbm \sigma_{{xx}} & \sigma_{{xy}} \\ \sigma_{{yx}} & \sigma_{{yy}}\ebm \tdot
\end{equation}
We define $\sigma$ as the square root of the greatest eigenvalue of the covariance matrix, \ie
\begin{equation}
\sigma^2 %
= \frac{\sigma_{{xx}} + \sigma_{{yy}}}{2} 
+ \frac{\sqrt{{4\sigma_{{xy}}^{2}} + \left(\sigma_{{xx}} - \sigma_{{yy}}\right)^{2}}}{2}
\tdot
\end{equation}
The estimation is robust to noise, works under asymmetrical distribution and is easy to use, but requires a parameter $\alpha$ to convert the standard deviation $\sigma$ into a pixel radius $\miradiuspix = \alpha\cdot\sigma$.
The parameter $\alpha$ is determined so that at least $98\%$ of the distribution is taken into account.
According to the standard normal distribution $Z$-score table, $\alpha$ is picked up in $\range{2.33, 2.37}$. In our experiments, we set $\alpha = 2.357$ as it best fits our measurements.

Recall that the pixel \ac{MI} radius is given by $\miradiuspix = \abs{\miradiusmm}/\pixelsize$.
The metric radius is either positive if formed after the rays inversion, as in \autoref{fig:blurradius}, or negative if before, and thus depends on the \textit{internal} configuration such as 
\begin{equation}\label{eq:miradiussgn}
\miradiusmm = \left\{
\begin{aligned}
 \miradiuspix \cdot \pixelsize &&\text{[Keplerian \textit{internal} configuration]}\tcomma\\
- \miradiuspix \cdot \pixelsize &&\text{[Galilean \textit{internal} configuration]}\tdot
\end{aligned}
\right.
\end{equation}

\paragraph{Coefficients estimation.}
Given several raw white images taken at different {apertures}, we estimate the parameters $\Omega$, \ie the coefficients of \autoref{eq:miradiusfnumber}, for each type of micro-image.
Note that the standard full-stop \tfnumber conventionally indicated on the lens differs from the real $f$-number.
We use then the $f$-number calculated from the {aperture} value $\aperturevalue$ by $\fnumber = \sqrt{2^{\aperturevalue}}$.
The coefficient $m$ is a function of fixed physical parameters independent of the micro-lens {focal lengths} and the main lens {aperture}.	
Therefore, we obtain a set of linear equations, sharing the same slope, but with different $y$-intercepts.
With $\vect{X} = \bbm m& q_1&\dots& q_\mltypenb\ebm\T$, the set of equations can be linearly rewritten as
$$\mat{A}\vect{X} = \vect{B}\tcomma \text{ and then } \vect{X} = \left(\mat{A}\T\mat{A}\right)^{-1}\mat{A}\T\vect{B}$$
where the matrix $\mat{A}$, containing the $f$-numbers and a selector of the corresponding $y$-intercept coefficient, and the vector $\vect{B}$, containing the radii measurements, are constructed by arranging the terms given the {focal length} at which they have been calculated.
Finally, we compute $\vect{X}$ with a least-square estimation.
\autoref{fig:radii} shows an example of radii distributions from our experiments computed from white images taken at several $f$-numbers, and the estimated linear functions.
In practice, at least two aperture configurations are required.
More can be used to improve the estimation but at the condition that radii measurement distributions are distinguishable from each others, with small overlap.

\subsection{Camera parameters initialization}\label{sec:paraminit}

First, the pixel size $\pixelsize$ is set according to the manufacturer values.
The main lens {focal length} $\Focal$ is also initialized from them. 
Given the parameters $\internalparam$ and the focus distance $\focusdist$, the parameters $\distmlasensor$ and $\distlensmla$ are initialized as
\begin{equation}\label{eq:initdd}
\distmlasensor \longleftarrow\frac{2m\Focusdist}{F + \xi\cdot 4m} \text{~~~~and~~~~} \distlensmla \longleftarrow \Focusdist - \xi\cdot 2\distmlasensor\tcomma
\end{equation}%
with $\xi = 1$ (\resp $\xi = -1$) in {Galilean} (\resp {Keplerian}) \textit{internal} configuration, and
where $\Focusdist$ is given by Eq.~(17) of \citet{Perwass2010c},
\begin{equation}\label{eq:focusdist}
\Focusdist = \abs{\frac{\focusdist}{2}\left(1-\sqrt{1 - 4\frac{\Focal}{\focusdist}}\right)}\tdot
\end{equation}
For completeness, note that the \textit{unfocused} configuration can be initialized with $\distmlasensor\leftarrow2m$ and $\distlensmla\leftarrow\Focal$.\\

In a second step, all distortions coefficients are set to zero.
The principal point is set as the center of the image.
The sensor plane is thus set parallel to the main lens plane, with no rotation, at a distance $- \left(\distlensmla+\distmlasensor\right)$.
Seemingly, the \ac{MLA} plane is initially set parallel to the main lens plane at a distance $-\distlensmla$. 
From the pre-computed \ac{MIA} parameters, the \ac{MLA} translation takes into account the $(x,y)$-offsets $\left(-\pixelsize\tau_x, -\pixelsize\tau_y\right)$ and the rotation around the $z$-axis is initialized with $-\vartheta_z$.
The micro-lenses pitch $\mlinterdist$ is set according to \autoref{eq:mlinterdist}, where the ratio $\lambda$ is computed using \autoref{eq:initdd} such as
\begin{equation}\label{eq:mldiameterapprox2}
\lambda \longleftarrow \frac{\Focal}{\Focal+2m}\tdot
\end{equation}
Finally, the initial micro-lenses' {focal lengths} are also computed from the parameters $\internalparam$ as follows
\begin{equation}
\focal\mltype{i} \longleftarrow \frac{\distmlasensor}{2 \cdot q_i'}\cdot\mlinterdist\tdot
\end{equation}
Experiments will show that the initial model is close to the optimized model.

\section{BAP features detection in raw images}\label{sec:detection}

At this point, the \ac{MIA} is calibrated and micro-images centers are extracted.
The raw images are devignetted by dividing them by a white raw image taken with the same {aperture}. 
We based our method on a checkerboard calibration pattern.
The detection process is divided into two steps:
1) checkerboard images are processed to extract corners at position $\left(u,v\right)$;
and 2) with the set of parameters $\internalparam$ and the associated virtual depth estimate for each corner, the corresponding \ac{BAP} feature is computed in image space.

\subsection{Computing blur radius through micro-lens}

To respect the $f$-number matching principle \citep{Perwass2010c}, we configure the main lens \tfnumber such that the micro-images fully tile the sensor without overlap.
In this configuration the working \tfnumber of the main imaging system and the micro-lens imaging system should match.
We consider the general case of measuring an object $\point$ at a distance $\distobj$ from the main lens.
First, $\point$ is projected through the main lens according to the thin lens equation,
${1}/{F} = {1}/{\distobj}+{1}/{\distimg}$, 
resulting in a point $\point'$ at a distance $\distimg$ behind the main lens, 
\ie at a distance 
$\distobj' =  \Dist - \distimg$
from the \ac{MLA}. 
From \autoref{eq:blurradius}, the metric radius of the blur circle $\blurradiusmm$ of a point $\point'$ at distance $\distobj'$ through a micro-lens of type $(i)$ is expressed as
\begin{align}\label{eq:radiusprefinal}
\blurradiusmm &= \left(\frac{\miinterdistmm\Dist}{\dist+\Dist}\right) \cdot \frac{\dist}{2} \cdot \left(\frac{1}{\focal\mltype{i}} - \frac{1}{\distobj'} - \frac{1}{\dist} \right) \nonumber\\
&= 
\underbrace{
	\frac{\miinterdistmm\cdot\Dist}{\dist+\Dist} \cdot \frac{\dist}{2} \cdot \frac{1}{\focal\mltype{i}}
}_{= q'_i \text{~[\autoref{eq:qprime}]}} 
-
\underbrace{
	\frac{\miinterdistmm\cdot\Dist}{\dist+\Dist} \cdot \frac{\dist}{2} \cdot \frac{1}{\dist}
}_{= \mlinterdist/2 \text{~[\autoref{eq:mlinterdist}]}}  - 
\underbrace{
	\frac{\miinterdistmm\cdot\Dist}{\dist+\Dist}
}_{= \mlinterdist \text{~[\autoref{eq:mlinterdist}]}} \cdot \frac{\dist}{2} \cdot \frac{1}{\distobj'}\nonumber\\
&= \left(-\mlinterdist\cdot\frac{\dist}{2}\right) \cdot \frac{1}{\distobj'} + \left( q_i'	- \frac{\mlinterdist}{2}\right)\tdot	
\end{align}
In practice, $\distobj'$ and $\dist$ cannot be measured in raw image space, but the virtual depth can, as it will be shown in the next subsection.
Virtual depth refers to relative depth value obtained from disparity. 
It is defined as the ratio between the signed object distance $\distobj'$ and the sensor distance $\dist$:
\begin{equation}
\virtdepth = -\frac{\distobj'}{\dist}\tdot
\end{equation}
The sign convention is reversed for virtual depth computation.
Distances are negative in front of the \ac{MLA} plane.
If we re-inject the virtual depth in \autoref{eq:radiusprefinal}, taking caution of the sign, and using \autoref{eq:mlinterdist}, we can derive the radius of the \textit{blur circle} of a point $\point'$ at a distance $\distobj'$ from the \ac{MLA} by
\begin{equation}\label{eq:radiusfinal}
\begin{aligned}
\blurradiusmm &= \frac{\mlinterdistapprox}{2}\cdot \virtdepth^{-1} + \left( q_i'	- \frac{\mlinterdistapprox}{2}\right)\tdot
\end{aligned}
\end{equation}
This equation allows to express the pixel radius of the blur circle $\blurradiuspix = \blurradiusmm/\pixelsize$ associated to each point having a virtual depth without explicitly evaluating the physical parameters $\aperture, \Dist, \dist, \Focal \tand \focal\mltype{i}$ of the camera, directly in image space.

\subsection{Features extraction}

First, we detect corners in raw images using the detector introduced by \citet{Noury2017b} with sub-pixel accuracy in each micro-image.
With a plenoptic camera, contrarily to a classic camera, a same point in object space is projected into multiple observations onto the sensor.
The checkerboard is designed and positioned so that the sets of observations are sufficiently far from each others to be clustered.
We use the \acs{DBSCAN} algorithm \citep{Ester1996} to identify the clusters.
We then associate each point with its cluster of observations.

Secondly, once each cluster is identified, we compute the virtual depth $\virtdepth$ from the disparity.
Let $\Delta\!\mlcenter_{1-2}$ be the distance between the centers of the micro-lenses $\mlcenter_1$ and $\mlcenter_2$, \ie the baseline.
Let $\Delta\!\point = \abs{\point_1 - \point_2}$ be the Euclidean distance between images of the same point in corresponding micro-images.
The virtual depth $\virtdepth$ is calculated with the intercept theorem:
\begin{equation}\label{eq:virtdepth}
\virtdepth = \frac{\Delta\!\mlcenter_{1-2}}{\Delta\!\mlcenter_{1-2} - \Delta\!\point} 
=  \frac{\eta\cdot\mlinterdist}{\eta\cdot\mlinterdist - \Delta\!\point} = \frac{\eta\cdot\mlinterdistapprox}{\eta\cdot\mlinterdistapprox - \Delta\!\point}\tdot
\end{equation}
If we consider two adjacent micro-lenses, the baseline $\Delta\!\mlcenter_{1-2}$ is just the diameter of a micro-lens, \ie $\mlinterdist = \mlinterdistapprox$ and $\eta = 1$.
For further apart micro-lenses the baseline is a multiple of that diameter, where $\eta$ is not necessarily an integer.
To handle noise in corner detection, we use a median estimator to compute the virtual depth of the cluster, taking into account all combinations of point pairs in the disparity estimation.

Finally, we compute the \ac{BAP} features from \autoref{eq:radiusfinal}, using the set of parameters $\internalparam$ and the available virtual depth $\virtdepth$.
In each frame $n$, for each micro-image $\left(\miindexes\right)$ of type $(i)$ containing a corner at position $\left(u,v\right)$ in the image, the feature $\observation$ is given by
\begin{equation}
\observation = \bbm u & v &\rho &1\ebm\T, \twith \rho = \blurradiusmm / \pixelsize\tdot
 \end{equation}
In the end, our observations are composed of a set of micro-images centers $\set{\micenter_{\miindexes}}$ and a set of \ac{BAP} features $\set{\observation}$ allowing us to introduce two reprojection error functions corresponding to each set of features as explains in the next section.

\section{Camera and relative blur calibration}\label{sec:calibration}

To retrieve the parameters of our camera model (\autoref{eq:completemodel}), 
we use a calibration process based on non-linear minimization of reprojection errors.
The camera calibration process is divided into three phases:
1) the initial intrinsics are provided by the pre-calibration step;
2) the initial extrinsics are estimated from the raw checkerboard images;
and 3) the parameters are refined with a non-linear optimization leveraging our new \ac{BAP} features.
In parallel, using our \ac{BAP} features, the blur proportionality coefficient of \autoref{eq:sigma2rho} is calibrated, by minimizing the relative blur in a new reprojection error with a non-linear optimization.

\subsection{Camera model initialization}

\begin{figure}[]
	\centering
	\includegraphics[width=\linewidth]{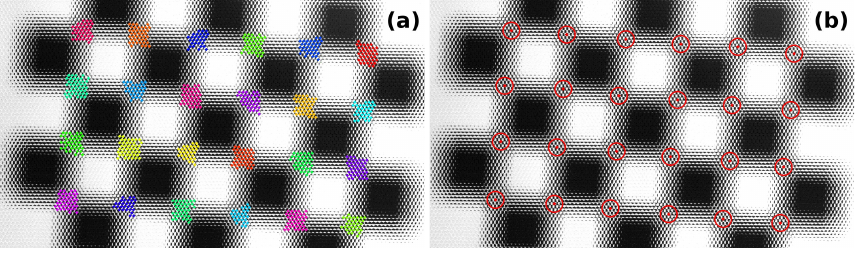}
	\caption{
		Checkerboard raw image with:
		\textbf{(a)} clusters of observations;
		\textbf{(b)} their barycenter used as approximation for extrinsics initialization.
	}\label{fig:cluster}
\end{figure}

Iterative optimization of non-linear cost functions are sensitive to initial parameters setting.
To ensure convergence and to avoid falling into local minima during the process, the parameters must be carefully initialized close to the solution.
Our pre-calibration step provides a strong initial solution for the optimization.
Intrinsic parameters are initialized as explained in \autoref{sec:paraminit} using only raw white images.

The camera poses $\set{\Transform\fcam^n}$, \ie the extrinsic parameters, are initialized using the same method as by \citet{Noury2017b}.
For each cluster of observations, the barycenter is computed, as illustrated by \autoref{fig:cluster}.
Those barycenters can been seen as the projections of the checkerboard corners through the main lens using a standard pinhole model.
For each frame, the pose is then estimated using the \ac{PnP} algorithm \citep{Kneip2011b}, like in classic pinhole imaging system.
To associate 3D-2D correspondences, we reproject checkerboard corners based on the estimated pose in image space and link them to their nearest cluster of observations.

\subsection{Optimizing the camera parameters}

By introducing blur in our model, we can optimize all parameters within one single optimization process.	
We propose a new cost function $\Theta$ taking into account the blur information of our new \ac{BAP} feature. 
The cost is composed of two main terms both expressing errors in the image space: 
1) the blur aware plenoptic reprojection error
and 2) the main lens center reprojection error.

In the first term, for each frame $n$, each checkerboard corner $\checkerboardcorner$ is reprojected into the image space through each micro-lens $\left(\mlindexes\right)$ of type $(i)$ according to the projection model of \autoref{eq:completemodel} and compared to its observations $\observation$.
In the second term, the main lens center $\lenscenter$ is reprojected according to a pinhole model in the image space through each micro-lens $\left(\mlindexes\right)$ and compared to its detected micro-image center $\micenter_{\mlindexes}$.
Let $\tens{S} = \set{\Xi, \set{\Transform\fcam^n}}$ be the set of intrinsic $\Xi$ and extrinsic $\set{\Transform\fcam^n}$ parameters to be optimized. 
The cost function $\Theta(\tens{S})$ is expressed as 
\begin{equation}
\Theta({\tens{S}})=%
\sum \norm{
	\observation - {\Proj_{\mlindexes}\left(\checkerboardcorner\right)}}^2 
+\sum \norm{\micenter_{\mlindexes} - \Proj_{\mlindexes}\left(\lenscenter\right)}^2\tdot
\end{equation}
The optimization is conducted using the Levenberg-Marquardt algorithm.

\subsection{Relative blur calibration using BAP features}

Relative blur estimation has been studied by \citet{Ens1993,Mannan2016b}.
Up to our knowledge, it has never been studied in context of plenoptic camera. 
As a new contribution, we leverage the relative blur between different micro-images and our \ac{BAP} features to calibrate the blur proportionality coefficient $\blurpropcoef$ of \autoref{eq:sigma2rho}.

\paragraph{Relative blur model.}
A point imaged by two different micro-lenses of type $(i)$ and $(j)$ will have different blur amount, \ie the resulting images will have different spread parameters for the \ac{PSF} model, such as
\begin{equation}\label{eq:imgformation}
\left\{
\begin{aligned}
\Img_{(i)}\!\left(x,y\right) &= \convp{\psf_{(i)}}{\Img^*\!\left(x,y\right)}\\
\Img_{(j)}\!\left(x,y\right) &= \convp{\psf_{(j)}}{\Img^*\!\left(x,y\right)}\tcomma
\end{aligned}
\right.
\end{equation}
where $\Img^*\!\left(x,y\right)$ is the latent in-focus image.
We approximate the \ac{PSF} with a 2D Gaussian as in \autoref{eq:psf}, where the diameter of the blur kernel $\psf_{(i)}$ is $\sigma_{(i)}$.
To compare two views with different amount of blur, we use the relative blur model in spatial domain \citep{Pentland1987,Subbarao1988,Subbarao1994,Ens1993}.
As stated by \citet{Mannan2016a}, the Gaussian relative blur approximation works well mainly for small relative blurs (up to $\blurradiuspix \approx 5~\si{pixels}$) and when the aperture has a simple shape, which is the case with the plenoptic camera.
We then use the equally-defocused representation by applying additional blur to the relatively in-focus micro-image, hence,
\begin{equation}\label{eq:relblurimgformation}
\left\{
\begin{aligned}
\Img_{(i)}\!\left(x,y\right) &\simeq \convp{\psf_{r}}{\Img_{(j)}\!\left(x,y\right)} &\text{if } \sigma_{(i)} \ge \sigma_{(j)} \\
\convp{\psf_{r}}{\Img_{(i)}\!\left(x,y\right) } &\simeq \Img_{(j)}\!\left(x,y\right) &\text{if } \sigma_{(i)} \leq \sigma_{(j)}\tdot\\
\end{aligned}
\right.
\end{equation}
Note that $\psf_{r}$ is the relative blur kernel applied to either one of the views such that both views are equally-defocused. 
The diameter of the relative blur kernel $\psf_{r}$ is approximated as
\begin{equation}
\sigma_r(i,j) \simeq \sqrt{|{\sigma_{(i)}^2 - \sigma_{(j)}^2}|}\tdot
\end{equation}
This approximation is exact when the \ac{PSF} is a Gaussian. 
Since the radius of the relative blur kernel $\sigma_r$ cannot indicate whether the $(i)$ or the $(j)$ view is more in-focus than the other, we define the relative blur similarly to \citet{Chen2015}, as 
\begin{equation}\label{eq:relblur}
\Delta\!\sigma^2(i,j) \triangleq {\sigma_{(i)}^2 - \sigma_{(j)}^2 },
\end{equation}
where $\Delta\!\sigma^2(i,j) > 0$ indicates that a pixel in the $(j)$-micro-image is more in-focus than its corresponding pixel in the $(i)$-micro-image.
Symmetrically, $\Delta\!\sigma^2(i,j) < 0$ indicates that the $(i)$-micro-image is more in-focus.
In a similar fashion, we define the {relative blur radius} as
\begin{equation}\label{eq:relblurradius}
\rho_r(i,j) \simeq \sqrt{\abs{\Delta\!\rho^2(i,j)}} = \sqrt{|{\rho_{(i)}^2 - \rho_{(j)}^2}|}
\end{equation}
with $\sigma_r = \kappa \cdot {\rho_r}$, and where $\rho_{(i)}, \rho_{(j)}$ are the blur radii of the \ac{BAP} features through a micro-lens of type $(i)$ and $(j)$.

\paragraph{Blur proportionality coefficient calibration.}

To calibrate $\blurpropcoef$, we use our \ac{BAP} features and the relative blur model applied on micro-images of different types.
\ac{BAP} features $\set{\feature_i}$ from a same cluster $\tens{C}$ represent the same point in object space $\point{}\fworld$.
We extract two windows $\tens{W}$ around the \ac{BAP} features $\feature{}_{i}, \feature{}_{j} \in \tens{C}\!\left(\point{}\fworld\right)$ of different types, 
and express them using the {equally-defocused} representation (\autoref{eq:relblurimgformation}).
As the relative blur radius does not exceed \SI{2.5}{pix}, windows $\tens{W}$ of size $9 \times 9$ are extracted at $\left(u, v\right)$ with sub-pixel precision, and represent therefore the same part of the scene in both micro-images.
Additional blur is applied using a Gaussian kernel of spread parameter $\sigma_r$.
The spread parameter is computed from the $\rho$ part of the \ac{BAP} features and the parameter $\blurpropcoef$ to be optimized, with initial value ${\blurpropcoef} = 1$.
Let $\Theta(\kappa)$ be the cost function to be minimized. It is expressed as
\begin{equation}
\Theta(\kappa) = 
\sum_n \sum_{
	\feature{}^{n}_{i}, \feature{}^{n}_{j} \in \tens{C}\left(\point{}^{n}\fworld\right)
} 
\norm{
	\tens{W}\!\left(\feature{}^{n}_{j}\right) - \convp{\psf_{r}}{	\tens{W}\!\left(\feature{}^{n}_{i}\right)}
}_2^{2}\tcomma
\end{equation}
given $\abs{\rho_{(i)}} < \abs{\rho_{(j)}}$ 
and where $\psf_{r}$ is the \ac{PSF} with spread parameter $\sigma_r = \blurpropcoef \cdot \sqrt{|{\rho_{(i)}^2 - \rho_{(j)}^2}|}$.
The optimization is conducted using the Levenberg-Marquardt algorithm.

\section{Experimental setup}\label{sec:setup}

\begin{figure*}[]
\centering
	\includegraphics[width=\linewidth]{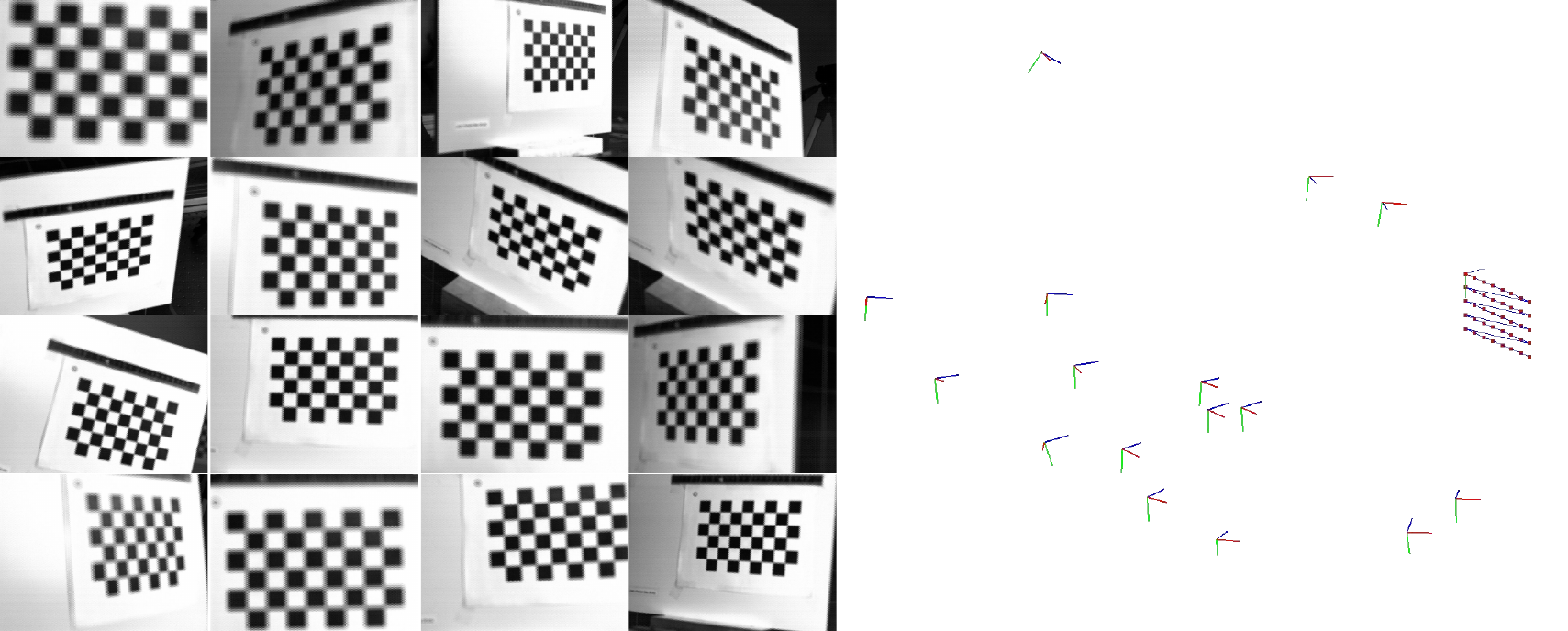}%
	\caption{
		Example of calibration targets acquired for distances between $775$ and $400$ \si{\mm} from the checkerboard used in the dataset \texttt{R12-B}, and their respective poses in 3D.
	}\label{fig:datasetr12b}
\end{figure*}

To validate our camera model, we evaluate our method on real-world data obtained with a multi-focus plenoptic camera in a controlled environment. 
Our experimental setup is illustrated in \autoref{fig:mfpc}.
The camera is mounted on a linear motion table with micro-metric precision.
The target plane is orthogonal to the translation axis, and the camera optical axis is aligned with this axis.
The approximate absolute distances at which the images have been taken with the corresponding step lengths are reported in \autoref{tab:datasets}.

\subsection{Hardware environment}

For our experiments we used a \texttt{Raytrix R12} color 3D-light-field-camera, with a \ac{MLA} of F/2.4 {aperture}.
The camera is in Galilean \textit{internal} configuration.
We used two different mounted lens, a \texttt{Nikon AF Nikkor F/1.8D} with a \SI{50}{\milli\meter} {focal length} {for comparison with state-of-the-art}, and a \texttt{Nikon AF DC-Nikkor F/2D} with a \SI{135}{\milli\meter} {focal length} {to validate the generalization of our model}.
The \ac{MLA} organization is hexagonal row-aligned, and composed of $176\times152$ (width $\times$ height) micro-lenses with $\mltypenb=3$ different types.
The sensor is a \texttt{Basler beA4000-62KC} with a pixel size of $\pixelsize= 0.0055$ \si{\milli\meter}.
The raw image resolution is $4080\times3068~\si{pixel}$.
We calibrate our camera for four focus distance configurations, with $\focusdist \in \set{450, 1000, \infty}$ \si{\mm} for the \SI{50}{\mm} lens, and with $\focusdist = 1500~\si{\mm}$ for the \SI{135}{\mm} lens.
Note that when changing the focus setting, the main lens moves with respect to the block \ac{MLA}-sensor.

\subsection{Software environment}

All images have been acquired using the \texttt{MultiCamStudio} free software (v6.15.1.3573) of the \texttt{Euresys} company.
We set the shutter speed to $5$ \si{ms}.
While taking white images for the pre-calibration step, we set the gain to its maximum value. %
For \texttt{Raytrix} data, we use their proprietary software \texttt{RxLive} (v4.0.50.2) to calibrate the camera, and compute the depth maps used in the evaluation.
Our source code has been made publicly available: \url{https://github.com/comsee-research/libpleno}, and
\url{https://github.com/comsee-research/compote}.
\begin{table}[b]
	\caption{
		Summary of \texttt{R12-A,B,C,D}, and \texttt{UPC-S} datasets contents. All distances are given in \si{\mm}. Scale refers to checkerboard square size. Evaluation distances refer to the linear motion table setup.
	}\label{tab:datasets}
	\begin{tabu}{X l cc cc ccc}	
		\toprule
		&& \multicolumn{2}{c}{Target} & \multicolumn{2}{r}{Calib. dist.} & \multicolumn{3}{c}{Eval. dist.} \\ 
		\cmidrule{3-4}\cmidrule(lr){5-6}\cmidrule{7-9}
		&$\focusdist$ & size & scale & min & max & min & max & step \\
		\midrule
		\midrule
		\texttt{A} &450 & $9\times5$ & 10 & 175 & 400 & 265 & 385 & 10\\
		\texttt{B} &1000 & $8\times5$ & 20 & 400 & 775 & 450 & 900 & 50\\
		\texttt{C} &$\infty$ & $6\times4$ & 30 & 500 & 2500 & 400 & 1250 & 50\\
		\midrule
		\texttt{D} &1500 & $5\times 3$ & 20 & 850 & 1300 & 750 & 1200 & 50\\
		\midrule
		\texttt{S} &hyperf. & $9\times 6$ & 26.25 & 250 & 800 & 200 & 500 & 50\\
		
		\bottomrule
	\end{tabu}
\end{table}

\subsection{Datasets}

We build four datasets with different focus distance $\focusdist$: 
for the \SI{50}{\mm} lens, 
\texttt{R12-A} for $\focusdist = 450~\si{\mm}$,
\texttt{R12-B} for $\focusdist = 1000~\si{\mm}$, 
and \texttt{R12-C} for $\focusdist = \infty$; 
for the \SI{135}{\mm} lens, \texttt{R12-D} for $\focusdist = 1500~\si{\mm}$.
Each dataset is composed of:
\begin{itemize}
	\item white raw plenoptic images acquired at different {apertures} ($\fnumber \in \set{4, 5.66, 8, 11.31, 16}$) using a light diffuser mounted on the main objective for pre-calibration,
	\item free-hand calibration target images acquired at various poses (in distance and orientation), separated into two subsets, one for the calibration process (\num{16} images) and the other for reprojection error evaluation (\num{15} images),
	\item a white raw plenoptic image acquired in the same luminosity condition and with the same {aperture} as in the calibration targets acquisition for devignetting,
	\item and, calibration targets acquired with a controlled translation motion for quantitative evaluation, along with the depth maps computed by the \texttt{RxLive} software.%
\end{itemize}
Examples of calibration targets acquired for the \texttt{R12-B} dataset are given in \autoref{fig:datasetr12b} along with their 3D poses. %
A summary for each dataset is given in \autoref{tab:datasets}, indicating  checkerboard information and the distances at which the targets have been acquired for calibration and for the controlled evaluation.
Our datasets have been made publicly available, and can be downloaded from our public repository at \url{https://github.com/comsee-research/plenoptic-datasets}.

\subsection{Simulation environment}

In order to validate our model on \texttt{Lytro}-like plenoptic camera configuration, \ie \textit{unfocused} plenoptic camera (UPC), we propose to evaluate our model in a simulation environment.
We built our own simulator based on raytracing to generate images.
Similar to the real-world dataset, we generated a dataset, named \texttt{UPC-S}, composed of several white images taken at different apertures (with $N \in \set{2, 4, 5.6}$), various checkerboard poses for calibration and validation, and for evaluation, checkerboard images with known translation along the $z$-axis.
Details are also given in \autoref{tab:datasets}.
We used the \texttt{Lytro Illum} intrinsic parameters reported in Table~4 of \citet{Bok2017} as baseline for the simulation.
They have been converted into our parameters and reported in \autoref{tab:intrinsicsimu}.
The \ac{MLA} organization is hexagonal row-aligned, and composed of $541\times 434$ (width $\times$ height) micro-lenses of the same type ($\mltypenb = 1$).
The raw image resolution is $7728\times 5368~\si{pixel}$, with a pixel size of $\pixelsize= 0.0014~\si{\milli\meter}$ and with micro-image of radius \SI{7.172}{pixel}. 

\section{Results and Discussions}\label{sec:results}

Our evaluation process follows the steps given in the overview (\autoref{fig:overview}).
First, we present the pre-calibration results, where white raw plenoptic images are used for computing micro-image centers, and for estimating initial camera parameters. 
Second, from the set of devignetted calibration target images, \ac{BAP} features are extracted, 
and camera intrinsic and extrinsic parameters are then computed using our non-linear optimization process.
In parallel, the same \ac{BAP} features are also used to calibrate the relative blur proportionality coefficient.
Third, we evaluate our model quantitatively, firstly, using the reprojection error as a metric, and secondly, using the relative translation error in a controlled environment.
Then, we propose an ablation study of the camera parameters.
Finally, we illustrate how to characterize the plenoptic camera extended \ac{DoF} using the blur profile.

\subsection{Pre-calibration}

To estimate the parameters $\internalparam$, we set $\alpha = 2.357$, and since the camera is in Galilean \textit{internal} configuration, we use $\miradiusmm = - \miradiuspix \cdot \pixelsize$,~ following \autoref{eq:miradiussgn}.
\autoref{fig:radii} shows the micro-image radii as function of the inverse $f$-number with the estimated lines for dataset \texttt{R12-B}.
Their distributions are represented by the violin-boxes.
For $N=5.66$, we can see that radii distributions overlap, and that radii values are slightly overestimated as they do not fit exactly the borders of the micro-images.
In practice, we only use white images that present distinguishable radii distributions in the estimation process, usually corresponding to small apertures.
In case of \texttt{R12-B}, only white images at $N=11.31$ and $N=8$ are used.
The corresponding coefficients for all datasets are summarized in \autoref{tab:internals}.
As expected, the parameter $m$ is different for each dataset, since $\Dist$ and $\miinterdistmm$ vary with the focus distance $\focusdist$, whereas
the $q'_i$ values are close for all datasets,
{even for different camera setup (\texttt{R12-D}).}
 
\begin{table}[]
	\caption{
		Set of parameters $\internalparam$ (in \si{\micro\meter}) computed during the pre-calibration step for each dataset, along with the calibrated relative blur proportionality coefficient.
	}\label{tab:internals}
	\begin{tabu}{c X X X X}
		\toprule
		& {\texttt{R12-A}} & {\texttt{R12-B}}  & {\texttt{R12-C}} & {\texttt{R12-D}}\\
		\cmidrule(r){2-4}\cmidrule(l){5-5}
		\midrule
		$\miinterdistmm$ &\nump{3}{128.22163189288707} &\nump{3}{128.29345695445207} &\nump{3}{128.333} & \nump{3}{127.8514120625554}\\
		$\lambda$ &{0.99441} &{0.99358} &{0.99380}& {0.99746}\\
		\midrule
		$m$&\nump{3}{-140.59554040431976} &\nump{3}{-159.56199169158936} &\nump{3}{-155.97544610500336}& \nump{3}{-171.28813340522536}\\
		$q'_1$&\nump{3}{35.135191355} &\nump{3}{036.489007285581368} &\nump{3}{035.442640815322377}& \nump{3}{038.5990534134653}\\
		$q'_2$&\nump{3}{40.268205527646997} &\nump{3}{042.07459207509591} &\nump{3}{041.277697120254017}&\nump{3}{043.129202376221407}\\
		$q'_3$&\nump{3}{36.822146226151445} &\nump{3}{038.806929475413102} &\nump{3}{037.858320210521199}&\nump{3}{040.787991420297272}\\
		\midrule
		$\blurpropcoef$ &\nump{4}{0.69883150043087894} &\nump{4}{0.69894622350796487} &\nump{4}{0.6531217254542494}& \nump{4}{0.88244764614033244}\\
		\bottomrule
	\end{tabu}
\end{table}

\subsection{Free-hand camera calibration}

\begin{table*}[]
	\caption{
		Initial intrinsic parameters for each dataset along with the optimized parameters obtained by our method (\texttt{BAP}) and with the methods of \citet{Noury2017b} (\texttt{NOUR}), of \citet{Nousias2017} for each micro-lens type (\texttt{NOUS1, NOUS2, NOUS3}) and the parameters obtained from \texttt{RxLive} software (\texttt{RTRX}).
		Reference and initial intrinsic parameters for the simulated \texttt{Lytro} dataset \texttt{UPC-S} along with the optimized parameters obtained by our method (\texttt{BAP}).
	}\label{tab:intrinsics}\label{tab:initials}\label{tab:intrinsicsimu}
	\begin{center}\footnotesize
		\begin{tabu}{r XXXXXXX XXXXXXX}
			\toprule
			&\multicolumn{7}{c}{\texttt{R12-A}~ ($\Focal = 50~\si{\mm}\tcomma\focusdist=450~\si{\mm}$)} & \multicolumn{7}{c}{\texttt{R12-B}~ ($\Focal = 50~\si{\mm}\tcomma\focusdist=1000~\si{\mm}$)}\\
			\cmidrule(r){2-8}\cmidrule(l){9-15}
			& Init. & \texttt{BAP} & \texttt{NOUR} & \texttt{NOUS1} & \texttt{NOUS2} & \texttt{NOUS3} & \texttt{RTRX}  
			& Init. & \texttt{BAP} & \texttt{NOUR} & \texttt{NOUS1} & \texttt{NOUS2} & \texttt{NOUS3} & \texttt{RTRX}\\
			\midrule
			\midrule
			$\Focal$ [\si{\mm}]&
			\num{50} &\nump{3}{49.714480}  &\nump{3}{54.8879866} &\nump{3}{61.3053547} &\nump{3}{62.4764324} &\nump{3}{63.32832} &\nump{3}{47.709} & %
			\num{50} &\nump{3}{50.0471219361} &\nump{3}{51.26219850} &\nump{3}{53.9128624729}  &\nump{3}{52.9875487405212} &\nump{3}{52.97728194}&\nump{3}{50.8942}\\ %
			$Q_1$ [$\times10^{-5}$]&
			0 &\nump{2}{24.65725}  &\nump{3}{6.098537} &- &- &- &- &  %
			0 &\nump{3}{2.89951703} &\nump{3}{0.022713011013}&- &- &- &- \\ %
			$-Q_2$ [$\times10^{-6}$]&
			0 &\nump{3}{2.99839}  &\nump{3}{0.925168}&- &- &- &- &  %
			0 &\nump{3}{0.299757477} &\nump{3}{0.09300222}&- &- &- &- \\%OK
			$Q_3$ [$\times10^{-8}$]&
			0 &\nump{3}{1.0628}  &\nump{3}{0.3033027}&- &- &- &- &  %
			0 &\nump{3}{0.0635259837} &\nump{3}{0.0065043984}&- &- &- &-\\ %
			$P_1$ [$\times10^{-5}$]&
			0 &\nump{1}{-14.58525}  &\nump{1}{-15.02797}&- &- &- &- &  %
			0 &\nump{2}{14.133984070} &\nump{2}{12.142952}&- &- &- &- \\%OK
			$-P_2$ [$\times10^{-5}$]&
			0 &\nump{3}{6.340192}  &\nump{3}{5.01985}&- &- &- &- &  %
			0 &\nump{2}{21.538891172} &\nump{2}{18.164420950336738}&- &- &- &- \\%OK
			\midrule
			$\Dist$ [\si{\mm}] &
			\nump{3}{56.6576} &\nump{3}{56.700741} &\nump{3}{62.4249} &\nump{3}{71.13062} &\nump{3}{72.54080} &\nump{3}{73.53018}& -&  %
			\nump{3}{52.1127} &\nump{3}{52.124834510} &\nump{3}{53.296254614647317}& \nump{3}{56.06204565} &\nump{3}{55.1279714770} &\nump{3}{55.1243615790}&-\\ %
			$-t_x$ [\si{\mm}] &
			\nump{2}{11.2926} &\nump{2}{10.9743487} &\nump{3}{9.771212}&- &- &- &- & %
			\nump{2}{11.2987} &\nump{2}{12.4383563} &\nump{2}{12.671744569515548}&- &- &- &-\\%OK
			$-t_y$ [\si{\mm}] &
			\nump{3}{8.4106} &\nump{3}{7.887169} &\nump{3}{8.33433107}&- &- &- &- &  %
			\nump{3}{8.41603} &\nump{3}{5.988136336} &\nump{3}{6.1139257658743213}&- &- &- &- \\%OK
			$-\theta_x$ [\si{\micro\radian}] &
			0 &\nump{1}{843.1} &\nump{1}{468.6}&- &- &- &- &
			0 &\nump{1}{607.2} &\nump{1}{576.2}&- &- &- &- \\
			$\theta_y$ [\si{\micro\radian}] &
			0 &\nump{1}{637.1} &\nump{1}{321.8}&- &- &- &- &
			0 &\nump{1}{514.5} &\nump{1}{350.4}&- &- &- &-\\
			$\theta_z$ [\si{\micro\radian}] &
			\nump{1}{0.552769601998199} &\nump{1}{31.5} &\nump{1}{25.3}&- &- &- & \nump{1}{41.87424}&  
			\nump{1}{016.9890424352331} &\nump{1}{46.0} &\nump{1}{35.3}&- &- &- & \nump{1}{41.87424}\\
			$\mlinterdist$ [\si{\micro\m}] &
			\nump{2}{127.505} &\nump{2}{127.45529} &\nump{2}{127.38116}&- &- &- & \nump{2}{127.357684}&  %
			\nump{2}{127.47} &\nump{2}{127.454102} &\nump{2}{127.40347541080915}&- &- &- & \nump{2}{127.357684}\\
			\midrule
			$\focal\mltype{1}$ [\si{\micro\m}]&
			\nump{2}{578.154} &\nump{2}{578.1820898851387} &-&-&-&-&-&  %
			\nump{2}{581.102} &\nump{2}{580.489071} &-&-&-&-&-\\
			$\focal\mltype{2}$ [\si{\micro\m}]&
			\nump{2}{504.456} &\nump{2}{505.41641184977304} &-&-&-&-&-&  %
			\nump{2}{503.958} &\nump{2}{504.31477775233} &-&-&-&-&-\\
			$\focal\mltype{3}$ [\si{\micro\m}]&
			\nump{2}{551.667} &\nump{2}{552.079041044199} &-&-&-&-&-&  %
			\nump{2}{546.393} &\nump{2}{546.3569893828} &-&-&-&-&-\\
			\midrule
			$\principalpointx$ [\si{pix}]&
			$2039$ & $2070.9$ &$2289.8$ &$1984.9$ &$2034.5$ &$1973.7$ &- & %
			$2039$ & $1958.3$ & $1934.9$ & $2074.7$ &$2094.7$ &$1837.0$ &- \\
			$\principalpointy$ [\si{pix}]&
			$1533$ & $1610.9$ &$1528.2$ &$1482.1$ &$1481.0$ &$1495.2$ &- & %
			$1533$ & $1802.9$ &$1759.3$ &$1640.2$ &$1649.1$ &$1620.4$ &- \\ %
			$\distmlasensor$ [\si{\micro\m}]&
			\nump{2}{318.632} &\nump{2}{324.774361}  &\nump{2}{402.323545}&- &- &- &- &  %
			\nump{2}{336.842} &\nump{2}{336.384159}  &\nump{2}{363.173728}&- &- &- &- \\ %
			\midrule
			$\distmlasensor\mltype{1}$ [\si{\micro\m}]&
			- &- &- &\nump{2}{585.15593609} &- &- &\nump{2}{407.806128263474} &  %
			- &- &- &\nump{2}{447.809603} &- &- &\nump{2}{407.806128263474} \\ %
			$\distmlasensor\mltype{2}$ [\si{\micro\m}]&
			- &- &- &- &\nump{2}{527.5860328} &- &\nump{2}{405.99712729454} &  %
			- &- &- &- &\nump{2}{401.9258803} &- &\nump{2}{405.99712729454} \\ %
			$\distmlasensor\mltype{3}$ [\si{\micro\m}]&
			- &- &- &-&-&\nump{2}{561.9345080977002} &\nump{2}{406.895726919174} &  %
			- &- &-  &- &- &\nump{2}{414.3172396} &\nump{2}{406.895726919174}\\ %
			\bottomrule
		\end{tabu}%
		\vspace*{1cm}\vfill
		\begin{tabu}{r XXXXXXX XX XXX} %
			\toprule
			&\multicolumn{7}{c}{\texttt{R12-C}~ ($\Focal = 50~\si{\mm}\tcomma\focusdist=\infty$)} & \multicolumn{2}{c}{\texttt{R12-D}} & \multicolumn{3}{c}{\texttt{UPC-S}}\\ %
			\cmidrule(r){2-8}\cmidrule(lr){9-10}\cmidrule(l){11-13}
			& Init. & \texttt{BAP} & \texttt{NOUR} & \texttt{NOUS1} & \texttt{NOUS2} & \texttt{NOUS3} & \texttt{RTRX}
			& Init. & \texttt{BAP} & Ref. & Init. & \texttt{BAP} \\% & \texttt{NOUR} & \texttt{NOUS1} & \texttt{NOUS2} & \texttt{NOUS3} & \texttt{RTRX}  \\
			\midrule
			\midrule
			$\Focal$ [\si{\mm}]&
			\num{50} &\nump{3}{50.013031225428279} &\nump{3}{53.321663056716744}& \nump{3}{51.11312043379682}&\nump{3}{49.91884989211353} &\nump{3}{50.81240303967532}  &\nump{3}{51.5635} &  %
			\num{135} &\nump{3}{136.10516670930977}&\nump{4}{9.98445} &\nump{3}{10}  &\nump{3}{10.229728755768972}\\%&\nump{3}{131.74702659856177} &\nump{3}{135}  &\nump{3}{135} &\nump{3}{135}&\nump{3}{70.0492}\\ %
			$Q_1$ [$\times10^{-5}$]&
			0 &\nump{2}{18.6134905} &\nump{3}{1.3927179538351636} &- &- &- &- &%
			0 &\nump{3}{35.974498894915505} & 0 & 0 & -\nump{3}{7.9132878299} \\%&\nump{3}{-2.5241223785800029}&- &- &- &- \\%OK
			$-Q_2$ [$\times10^{-6}$]&
			0 &\nump{3}{2.64623248} &\nump{3}{0.3820504079168516} &- &- &- &-&%
			0 &\nump{3}{8.083212390028945} & 0 & 0 & -\nump{3}{3.8819307}   \\%&\nump{3}{-0.058224366746762891}&- &- &- &-  \\%OK
			$Q_3$ [$\times10^{-8}$]&
			0 &\nump{3}{1.0376424096} &\nump{3}{0.10405646336111867} &- &- &- &-& %
			0 &\nump{3}{4.820932981690617}  & 0 & 0 & -\nump{3}{5.8213744} \\% &\nump{3}{-0.005975996247027278}&- &- &- &- \\%OK
			$P_1$ [$\times10^{-5}$]&
			0 &\nump{2}{19.10910136779} &\nump{2}{27.719919106678484} &- &- &- &-&%
			0 &\nump{2}{-4.3116385625392565}  & 0 & 0 & -\nump{3}{3.5767} \\% &\nump{2}{5.2361435610774322}&- &- &- &- \\%OK
			$-P_2$ [$\times10^{-5}$]&
			0 &\nump{3}{7.31057313775} &\nump{3}{4.4607044103427373} &- &- &- &-&%
			0 &\nump{2}{-3.7582001008085353} & 0 & 0 & \nump{3}{0.116994898} \\%&\nump{2}{6.0596357146373565}&- &- &- &- \\%OK
			\midrule
			$\Dist$ [\si{\mm}] &
			\nump{3}{49.3841} &\nump{3}{49.3616079947} &\nump{3}{52.37898088844485} &\nump{3}{50.33135166338584} &\nump{3}{49.06695936005554} &\nump{3}{49.88220154361741}  &-&%
			\nump{2}{149.242563551545} &\nump{2}{149.10460493842936} &\nump{4}{9.847916429} &\nump{3}{10} &\nump{3}{10.005486613307905}\\% &\nump{2}{143.92420929262224}& \nump{3}{56.06204565} &\nump{3}{55.1279714770} &\nump{3}{55.1243615790}&-\\ %
			$-t_x$ [\si{\mm}] &
			\nump{2}{11.261763581600288} &\nump{2}{13.12502231402} &\nump{2}{14.158858382582752} &- &- &- &- &%
			\nump{2}{11.2987} &\nump{2}{11.208316218108749} & \nump{3}{5.45928} & \nump{3}{5.4047000000000001} & 0 \\%&\nump{2}{22.215363154147951}&- &- &- &- \\ %
			$-t_y$ [\si{\mm}] &
			\nump{3}{8.41781} &\nump{3}{7.445715535802826} &\nump{3}{6.2308459789508488} &- &- &- &- &%
			\nump{3}{8.3865828456157647} &\nump{3}{8.3508736830892687} & \nump{3}{3.7179} & \nump{3}{3.7478000000000002} & 0 \\% &\nump{2}{-0.64027490567511558}&- &- &- &- \\ %
			$-\theta_x$ [\si{\micro\radian}] &
			0 &\nump{1}{490.9} &\nump{1}{487.7}&- &- &- &- &
			0 &\nump{1}{371.11947118520142} & 0 & 0 & -\nump{3}{0.5295526} \\%&\nump{1}{147.58623470067671}&- &- &- &- \\
			$\theta_y$ [\si{\micro\radian}] &
			0 &\nump{1}{388.9} &\nump{1}{366.1}&- &- &- &- &
			0 &\nump{1}{287.0474135919366} & 0 & 0 &  \nump{3}{0.94384387} \\% &\nump{1}{221.68098150729369}&- &- &- &- \\
			$\theta_z$ [\si{\micro\radian}] &
			\nump{1}{027.9612148213365} &\nump{1}{41.1} &\nump{1}{49.1} &- &- &- & \nump{1}{36.6290384017} &
			\nump{1}{5.794623978840437} &\nump{1}{35.445740626310149}  & 0 & 0 &  -\nump{3}{0.0332672775645}\\% &\nump{1}{48.546427050069222}&- &- &- & \nump{1}{31.6222482402} \\
			$\mlinterdist$ [\si{\micro\m}] &
			\nump{2}{127.537} &\nump{2}{127.48422481156} &\nump{2}{127.41756137206919} &- &- &- & \nump{2}{127.357684} &
			\nump{2}{127.52779727215502} &\nump{2}{127.5093511093891}  & \num{20} & \nump{3}{19.986916170727197} & \nump{3}{19.986878364538083}\\% &\nump{2}{127.477875302031}&- &- &- & \nump{2}{127.357684}\\ %
			\midrule
			$\focal\mltype{1}$ [\si{\micro\m}]&
			\nump{2}{554.348} &\nump{2}{569.878578597} &-&-&-&-&-&
			\nump{2}{625.62545776969802} &\nump{2}{636.06349373629145} &\nump{3}{40.087372} &\nump{3}{47.75307}  &\nump{3}{47.74745073675149}\\% &-&-&-&-&-\\ %
			$\focal\mltype{2}$ [\si{\micro\m}]&
			\nump{2}{475.984} &\nump{2}{491.71204034229} &-&-&-&-&-&
			\nump{2}{559.911826113209} &\nump{2}{572.51809391452524} & - &- &-\\% &-&-&-&-&-\\ %
			$\focal\mltype{3}$ [\si{\micro\m}]&
			\nump{2}{518.976} &\nump{2}{535.28388772589} &-&-&-&-&-&
			\nump{2}{592.05049379458463} &\nump{2}{604.22701441088145} & - &- &- \\% &-&-&-&-&-\\ %
			\midrule
			$\principalpointx$ [\si{pix}]&
			$2039$ &  $1692.1$ & $2131.6$&$1966.3$  &$1913.8$&$2052.5$ &-&
			$2039$ & $2028.7$ & $3842.8$ & $3863$ &$3861.7$\\% & $21.7$ & $2074.7$ &$2094.7$ &$1837.0$ &- \\ %
			$\principalpointy$ [\si{pix}]&
			$1533$ & $1677.8$ & $1445.9$  & $1484.6$& $1487.2$&$1492.7$ &- & %
			$1533$ & $1526.7$ &	$2719.5$ & $2683$ &$2715.5$\\% &$3165.7$ &$1640.2$ &$1649.1$ &$1620.4$ &- \\ %
			$\distmlasensor$ [\si{\micro\m}]&
			\nump{2}{307.929} &\nump{2}{319.527769} & \nump{2}{367.402085} &- &- &- &-& %
			\nump{2}{378.71822422746959} &\nump{2}{382.30433286906873} & \nump{3}{40.087372} &\nump{3}{47.2792}  &\nump{3}{47.322794127719447}\\ %
			\midrule
			$\distmlasensor\mltype{1}$ [\si{\micro\m}]&
			- &- &- &\nump{2}{357.7952714427} &- &- &\nump{2}{407.806128263474} & %
			- &- & - &- & -\\%&- &\nump{2}{447.809603} &- &- &\nump{2}{407.806128263474} \\  %
			$\distmlasensor\mltype{2}$ [\si{\micro\m}]&
			- &- &- &- &\nump{2}{349.9948302820} &- &\nump{2}{405.99712729454} & %
			- &- & - &- & -\\%&- &- &\nump{2}{401.9258803} &- &\nump{2}{405.99712729454} \\ %
			$\distmlasensor\mltype{3}$ [\si{\micro\m}]&
			- &- &-  &- &- &\nump{2}{353.26421436} &\nump{2}{406.895726919174}& %
			- &- & - &- & -\\%&-  &- &- &\nump{2}{414.3172396} &\nump{2}{406.895726919174}\\ %
			\bottomrule
		\end{tabu}%
		\vfill
	\end{center}	
\end{table*}

\paragraph{Comparison with state-of-the-art.}
Since our model is close to the one of~\citet{Noury2017b}, we compare our intrinsics with the ones obtained under their pinhole assumption using only corner reprojection error and with the same initial parameters.
In addition, we evaluate against the method of \citet{Nousias2017}, which provides a set of intrinsics and extrinsics for each micro-lens type. The equivalence of our parameters and their parameters is given by
\begin{equation}\label{eq:nousias}
\begin{aligned}
&&\Focal &= \frac{(f_x + f_y)}{2} \cdot \pixelsize\tcomma&&
\distlensmla = -\Focal\cdot \left(\frac{K_1}{K_2}\cdot\Focal + 1\right)^{-1}\tcomma\\
&&\distmlasensor\mltype{i} &= \distlensmla - \frac{K_2 \distlensmla}{\distlensmla + K_2}\tcomma&&
\principalpointx = c_x~~\tand~~\principalpointy = c_y\tcomma
\end{aligned}
\end{equation} 
where $K_1$ and $K_2$ are the two additional intrinsic parameters that account for the \ac{MLA} setting in their model.
The equivalence also stands for the parameters of \citet{Bok2017}.
The provided detector from \citet{Nousias2017} was not able to detect corner observations on our datasets. 
Therefore, we used the same observations for our method (noted \texttt{BAP} in \autoref{tab:intrinsics}), \citet{Noury2017b} method (\texttt{NOUR}), and \citet{Nousias2017} method for each type (\texttt{NOUS1}, \texttt{NOUS2}, and \texttt{NOUS3}), which allowed us to focus the comparison on the camera model only.
Finally, we provide the calibration parameters obtained from the \texttt{RxLive} software (\texttt{RTRX}) corresponding to the model of \citet{Heinze2016b}, and compare our depth measurements to their depth maps.

\paragraph{Initialization.}
We initialize $\lambda$ from \autoref{eq:mldiameterapprox2}. 
Its value for each dataset is reported in \autoref{tab:internals}.
The difference between the initial value of $\lambda$ and its value computed from optimized camera parameter is less than \SI{0.024}{\percent}, which validates the use of the initial value from \autoref{eq:mldiameterapprox2} when computing our \ac{BAP} features.
The initial camera parameters reported in \autoref{tab:initials} are computed using the methodology presented in \autoref{sec:paraminit}. %
They are used for the \texttt{BAP} and \texttt{NOUR} methods. %
The camera \textit{internal} configuration is set to Galilean.
When $\focusdist$ decreases, $\distlensmla$ increases.
Yet when the main lens focus distance is at infinity, the main lens should focus on the plane $\virtdepth = 2$, which implies that $\distlensmla$ tends to $\Focal - 2\distmlasensor$ as lower bound, as $\Focusdist$ tends to $\Focal$.
In most cases (here, for \texttt{R12-A,B,D}), we will still have $\Focal < \distlensmla$, which usually can describe the camera in Keplerian configuration.
In Keplerian \textit{internal} configuration, the condition $\Focal < \distlensmla$ stands regardless of the focus distance, as $\distlensmla$ lower bound is $\Focal + 2\distmlasensor$.
When using the linear initialization from \texttt{NOUS}, %
the initial parameters of some configurations corresponded to impossible physical setup or were too far from the solution, hindering the convergence of the optimization.
Therefore, in order to continue comparison, we manually set the initial parameters close enough to a solution. %
In contrast, we can see that the optimized parameters for \texttt{BAP} and \texttt{NOUR} are close to initial values, which shows that our pre-calibration step provides a strong initial solution for the optimization process.

\paragraph{Intrinsic camera parameters.} Optimized intrinsic parameters are reported for each dataset and for all the evaluated methods in \autoref{tab:intrinsics}.
First, \texttt{BAP}, \texttt{NOUR} and \texttt{NOUS} all verify the condition $\Focal \approx \distlensmla + 2\distmlasensor$ when the focus is set at infinity (\texttt{R12-C}).
Second, the focal lengths obtained from \texttt{NOUR}, \texttt{NOUS} and \texttt{RTRX} change significantly given the focus distance, and the ones obtained from \texttt{NOUS} even vary according to the micro-lens types.
In contrast, only \texttt{BAP} shows stable parameters across all three \texttt{R12-A,B,C} datasets. 
Shared parameters across datasets (\ie the focal lengths and the distance between the \ac{MLA} and the sensor) are close enough to indicate that our model successfully generalizes to different focus configurations.
{Furthermore, the parameters obtained by our method with an other main lens, \ie \texttt{R12-D}, are coherent with the previously obtained parameters, stressing out that our model can be applied to a different camera setting.}
Finally, our method is the only one providing the micro-lenses focal lengths in a single unified model. 
The other methods calibrate either several \ac{MLA}-sensor distances (\texttt{RTRX}), or several models, one for each type (\texttt{NOUS}).

Note that distortion coefficients and \ac{MLA} rotations are close to zero. 
The influence of these parameters will be analyzed in the proposed ablation study of the camera model in \autoref{sec:ablation}.

\paragraph{On simulated data.} 
First, pre-calibration has been performed using the white raw images.
The resulting parameters $\Omega$ are coherent with the simulation parameters.
With parameters $m = \SI{-23.639}{\micro\meter}$ and $q = \SI{-0.146}{\micro\meter}$, we have $\dist \approx \focal$, which describes the \textit{unfocused} configuration.
Reference and initial intrinsic parameters are reported in \autoref{tab:intrinsicsimu}, along with the optimized parameters.
Second, calibration has been performed. 
The obtained intrinsic parameters are close enough to the references parameters, indicating that our method is able to generalize to the \textit{unfocused} plenoptic camera.\\
For completeness, we also quantitatively evaluated the optimized parameters, by estimating the relative displacement between checkerboard with known motion along the $z$-axis.
It results a translation error $\varepsilon_z = \SI{1.64}{\percent}$, which validates the model.

\subsection{Quantitative evaluations of the camera model}

\paragraph{Reprojection error.}

In the absence of ground truth, we first evaluated the intrinsic parameters by estimating the reprojection error using the previously computed intrinsics.
We consider only free-hand calibration target images which are not used in the calibration process.
We use the \ac{RMSE} as a metric to evaluate the reprojection error on the corner part of the features, for each dataset.
For the \texttt{BAP} method, the corner reprojection part is reported in \autoref{tab:rmse}, as well as the radius reprojection part within parentheses.
Regarding the \texttt{NOUS} methods, the original error is expressed using the mean reprojection error (MRE). 
We converted the final error to the RMSE metric for comparison. 
Note that the latter method operates separately on each type of micro-lens, meaning that the number of features is not the same as with \texttt{NOUR} and \texttt{BAP}.
First, the reprojection error is less than \SI{1}{pixel} for all methods, for each dataset, demonstrating that the computed intrinsics lead to an accurate reprojection model and can be generalized to images which are not from the calibration set.
Second, even though the \texttt{NOUS} method provides the lowest RMSE, it shows a significant discrepancy according to the considered type. 
The error obtained by our method is sightly higher than the error from \texttt{NOUR}, but this can be explained by the fact that our optimization does not aim at minimizing only the corner reprojection error, but the radius reprojection as well.
Note that the positional error $\epsilon_{u,v}$ predominates in the total cost by two orders of magnitude compared to the blur radius error $\epsilon_\rho$, but the latter still helps to constrain our model as shown by the relatively close intrinsics between the datasets.

\paragraph{Controlled environment poses evaluation.}

With our experimental setup, we acquired several images with known relative translation between each frame.
We compare the estimated displacements along the $z$-axis from the extrinsic parameters to the ground truth.
The extrinsics are computed with the models estimated from the free-hand calibration.
In the case of the \texttt{RTRX} method, we use the filtered depth maps obtained with the proprietary software \texttt{RxLive} to estimate the displacements.
\begin{figure*}[!t]
\centering
		\includegraphics[width=\linewidth]{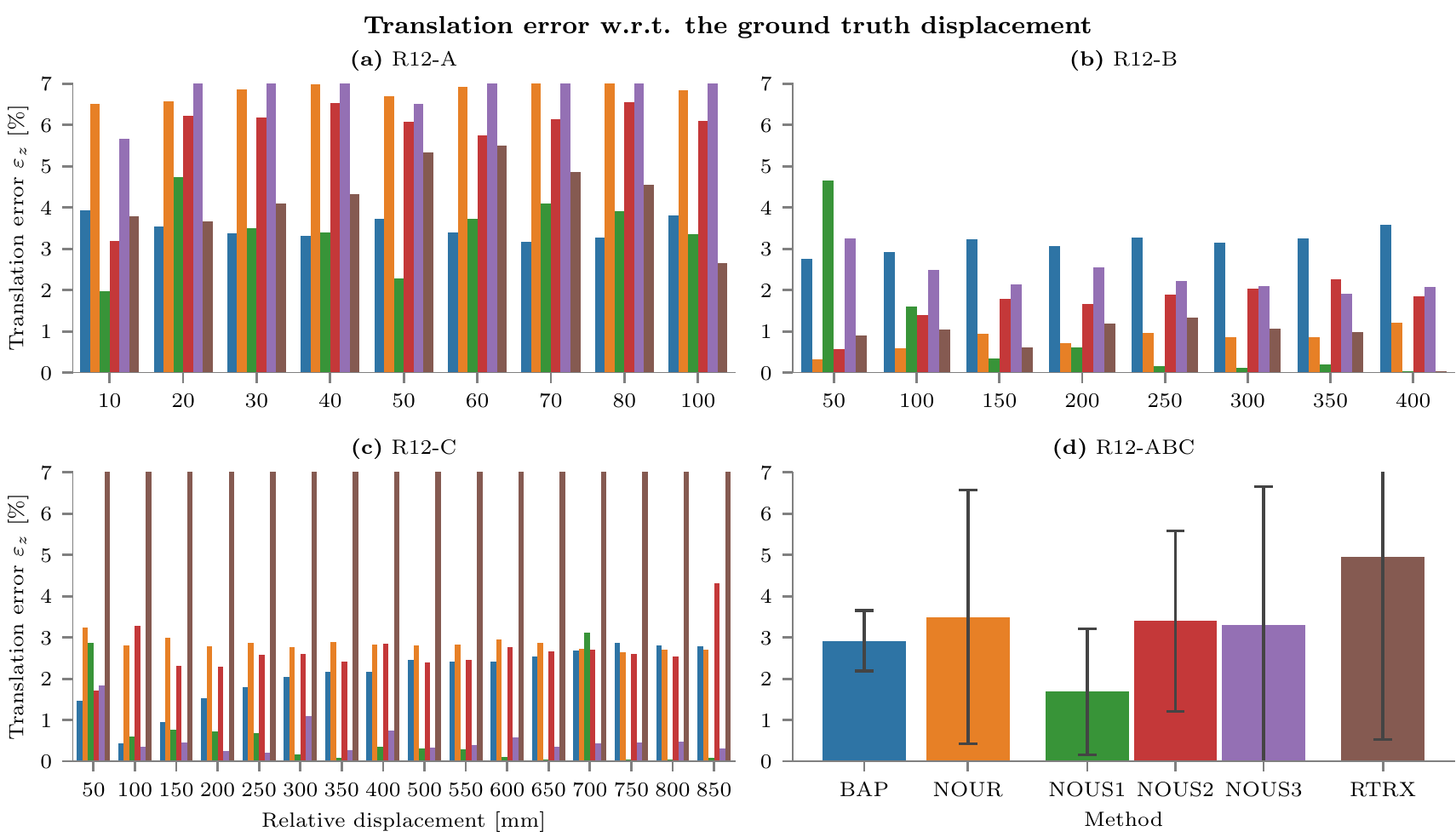}
	\caption{
		Translation error along the $z$-axis with respect to the ground truth displacement from the closest frame, for datasets \texttt{R12-A} \textbf{(a)}, \texttt{R12-B} \textbf{(b)} and \texttt{R12-C} \textbf{(c)}.
		The error $\varepsilon_z$ is expressed in percentage of the estimated distances, and truncated to \SI{7}{\percent} to ease the readability and the comparison.
		The mean error with its confidence interval across all datasets for our method (\texttt{BAP}), \citet{Noury2017b} method (\texttt{NOUR}), \citet{Nousias2017} method for each type (\texttt{NOUS1, NOUS2, NOUS3}), and for the proprietary software \texttt{RxLive} (\texttt{RTRX}) are reported in \textbf{(d)}.
		Please refer to the color version for better visualization.
	}\label{fig:translationerror}
\end{figure*}
The translation errors along the $z$-axis with respect to the ground truth displacement from the closest frame are reported in \autoref{fig:translationerror} for datasets \texttt{R12-A} {(a)}, \texttt{R12-B} {(b)} and \texttt{R12-C} {(c)}.
The relative error $\varepsilon_z$ for a known displacement $\delta_z$ is computed as the mean absolute relative difference between the estimated displacement $\hat{\delta_z}$ and the ground truth, for each pair of frames $\left(\Transform_i, \Transform_{j}\right)$ separated by a distance $\delta_z$, \ie
\begin{equation}
\funarg{\varepsilon_z}{\delta_z} = \eta^{-1} \sum_{\left(\Transform_i, \Transform_{j}\right) \mid z_i - z_j = \delta_z} {\abs{\delta_z - \hat{\delta_z}}} \textfractionsolidus {\delta_z}\tcomma
\end{equation}
where $\hat{\delta_z} = \hat{z_i} - \hat{z_j}$,
and $\eta$ is a normalization constant corresponding to the number of frames pair. %
The mean error with its standard deviation across all datasets for \texttt{BAP}, \texttt{NOUR}, \texttt{NOUS}, and \texttt{RTRX} are reported in {(d)}.

\begin{table}[!t]
	\caption{
		Corner reprojection error for each evaluation dataset (\ie free-hand calibration target images not part of the calibration dataset) using the \ac{RMSE} metric. For the \texttt{BAP} method, reprojection error of the radius part is indicated within parentheses.%
	}\label{tab:rmse}
	\begin{center}%
		\begin{tabu}{l c X XXX}
			\toprule
			& {\texttt{BAP}}& {\texttt{NOUR}} & {\texttt{NOUS1}} &{\texttt{NOUS2}} & {\texttt{NOUS3}}\\
			\midrule
			\midrule 
			\texttt{R12-A} & \nump{3}{0.856418} \scriptsize(\nump{3}{0.0826198}) & \nump{3}{0.712565}& \nump{3}{0.7729221250851441}& \textbf{{0.667}} %
			& \nump{3}{0.9582578903241914}\\
			\texttt{R12-B} & \nump{3}{0.673851} \scriptsize(\nump{3}{0.183383}) & \nump{3}{0.617605}&\nump{3}{0.5379751775899176}& \textbf{{0.519}}%
			& \nump{3}{0.5931335934878329}\\
			\texttt{R12-C} & \nump{3}{0.737945} \scriptsize(\nump{3}{0.0405954}) & \nump{3}{0.712565}&\nump{3}{1.287158262915446}&\nump{3}{0.6813681367003119}& \textbf{{0.411}} \\ %
			\bottomrule
		\end{tabu}
	\end{center}
\end{table}

Firstly, the mean error across \texttt{R12-A,B,C} datasets are of the same order for the evaluated methods around \SI{3}{\percent}: 
for \texttt{BAP}, $\varepsilon_z = 2.92 \pm 0.73~\si{\percent}$;
for \texttt{NOUR}, $\varepsilon_z = 3.50 \pm 3.08~\si{\percent}$;
for \texttt{NOUS1}, $\varepsilon_z = 1.68 \pm 1.53~\si{\percent}$;
for \texttt{NOUS2}, $\varepsilon_z = 3.40 \pm 2.19~\si{\percent}$;
for \texttt{NOUS3}, $\varepsilon_z = 3.30 \pm 3.35~\si{\percent}$; and,
for \texttt{RTRX}, $\varepsilon_z = 4.96 \pm 4.44~\si{\percent}$.
{This is also the case for the dataset \texttt{R12-D} where our model has a mean translation error of $\varepsilon_z = 3.37~\si{\percent}$.}
Note that all evaluated methods outperform \texttt{RTRX} as the depth maps computation might not be as precise as the optimization of extrinsic parameters. %
Our method ranks second in terms of relative mean error.
Even though lowest error is obtained by the method \texttt{NOUS} for type $(1)$, it presents a large standard deviation and the errors for the other two types are significantly higher.
In real application context, there is no way to know in advance which type will produce the smallest error.
\citet{Nousias2017} suggest that when extrinsics are sufficiently close, we can use representative extrinsics that are calculated by averaging the extrinsics from the individual types.
Our results do not match this observation as the estimated extrinsics are significantly different for each type.
As shown, only the first type gives satisfactory results whereas the other two present a larger error with a significant standard deviation.
Averaging the extrinsics from all types will therefore minimize the difference between poses but will not provide the best possible estimation.

Secondly, the standard deviation can be seen as an indicator of the estimation precision across the datasets, and thus indicates whether the model can generalize to several configurations or not.
Our model presents the lowest standard deviation as illustrated in \autoref{fig:translationerror} (d).
This indicates a low discrepancy between datasets and thus that the model is precise and consistent for all configurations.
Thirdly, we analyze the behavior of each method for each dataset across different distances.
None of the methods suffered from a constant bias, as we do not observe a decreasing relative error as the distance increases.
\texttt{BAP} and \texttt{NOUR} present a stable relative error for all distances, \ie with approximately \SI{0.3}{\percent} of standard deviation.
This indicates that the estimation suffered only from a scale error.
One could thus re-scale the poses to provide a precise and accurate estimation.
We cannot draw any conclusion for the other methods since the variations do not follow any obvious pattern.

Finally, our model differs from the model of \citet{Noury2017b} by modeling the micro-lens focal lengths.
Comparing those two models, the mean error as well as the standard deviation is smaller with our method. 
The inclusion of the micro-lens focal lengths in the camera model improves the estimation precision and accuracy, and enables to generalize to several configurations.
Dealing with different intrinsics which produce different extrinsics is not satisfactory when using the multi-focus plenoptic camera.
In contrast, our model is able to manage all micro-lens types simultaneously, and proves to be stable across various configurations and working distances.

\subsection{Ablation study of camera parameters}\label{sec:ablation}

To evaluate the influence of each parameter of the camera model, we present an ablation study of some of them.
We focus the analysis on distortion coefficients ($Q_1$, $Q_2$, $Q_3$, $P_1$, and $P_2$), 
on some degrees of freedom of the \ac{MLA}, especially its tilt with respect to the sensor ($\theta_x, \theta_y$), and the pitch between micro-lenses ($\mlinterdist$). %
All combinations of the parameters have been tested, resulting in eight configurations.
For each configuration and on each dataset of \texttt{R12-A,B,C}: first, we calibrate the camera intrinsic parameters; second, we evaluate the model using the \ac{RMSE} of the reprojection error; and finally, we quantitatively estimate the relative translation error on the evaluation dataset.
Each configuration has been initialized with the same intrinsic parameters, and used the same observations for all processes.
Results are reported in \autoref{tab:ablation}.
The first column is the configuration number.
The \texttt{Tilt} column indicates if we keep (\checkmark) or remove ($\times$) the parameters $\theta_x$ and $\theta_y$.
The \texttt{Pitch} column stands for the parameter $\mlinterdist$, and the column \texttt{Dist} for the distortion parameters $Q_1$, $Q_2$, $Q_3$, $P_1$, and $P_2$.
The reprojection error ${\epsilon_{all}}$ is given by its \ac{RMSE}, and the relative translation error ${\epsilon_z}$ is expressed in percent with respect to the ground truth displacement.
\begin{table}[]
	\caption{
		Ablation study of some camera parameters.
		For each dataset, the reprojection error ${\epsilon_{all}}$, computed using the \ac{RMSE} along with the relative translation error ${\epsilon_z}$, expressed in \si{\percent}, are reported.
		The symbol \checkmark (\resp $\times$) indicates if we keep (\resp remove) the considered parameters.%
	}\label{tab:ablation}
	\begin{tabu}{X XXX cc cc cc}
		\toprule
		&
		\multirow{2}{*}{\rotatebox[origin=l]{90}{\texttt{Tilt}}} &
		\multirow{2}{*}{\rotatebox[origin=l]{90}{\texttt{Pitch}}} &
		\multirow{2}{*}{\rotatebox[origin=l]{90}{\texttt{Dist}}} &
		\multicolumn{2}{c}{\texttt{R12-A}}& \multicolumn{2}{c}{\texttt{R12-B}} & \multicolumn{2}{c}{\texttt{R12-C}} \\
		\cmidrule{5-6}\cmidrule(lr){7-8}\cmidrule{9-10}
		&&&& ${\epsilon_{all}}$& ${\epsilon_z}$ & ${\epsilon_{all}}$ & ${\epsilon_z}$ & ${\epsilon_{all}}$ & ${\epsilon_z}$ \\
		\midrule
		\midrule 
		1& \checkmark & \checkmark & \checkmark &
		\textbf{0.860} & \textbf{3.23} & %
		\textbf{0.698} & \textbf{3.15} & %
		\nump{3}{0.738684029}&\nump{2}{2.31038113}\\%R12C
		\midrule
		2& \checkmark & \checkmark & $\times$ &
		\nump{3}{0.86642207}&\nump{2}{3.33891}& %
		\nump{3}{0.700050924}&\nump{2}{3.197026}& %
		\textbf{0.737} %
		&\nump{2}{3.18429257}\\%R12C
		3& \checkmark & $\times$ & \checkmark &
		\nump{3}{0.884110041}&\nump{2}{3.876323}& %
		\nump{3}{0.755160051}&\nump{2}{3.2118134226}& %
		\nump{3}{0.770359045}&\textbf{2.00}\\ %
		4& \checkmark & $\times$ & $\times$ &
		\nump{3}{0.891059207}&\nump{2}{3.976911865}& %
		\nump{3}{0.751962488}&\nump{2}{3.235505357}& %
		\nump{3}{0.77273193}&\nump{2}{3.133764226}\\%R12C
		5& $\times$ & \checkmark & \checkmark &
		\nump{3}{0.864695193}&\nump{2}{3.483333}& %
		\nump{3}{0.784339725}&\textbf{3.15} & %
		\nump{3}{0.759739712}&\nump{2}{2.894325984}\\%R12C
		6& $\times$ & \checkmark & $\times$ &
		\nump{3}{0.86357087}&\nump{2}{3.58049377}& %
		\nump{3}{0.7164409}&\nump{2}{3.15608631}& %
		\nump{3}{0.748509444}&\nump{2}{3.042542619}\\%R12C
		\midrule
		7& $\times$ & $\times$ & \checkmark &
		-&-& %
		-&-& %
		-&-\\%R12C
		8& $\times$ & $\times$ & $\times$ &
		-&-& %
		-&-& %
		-&-\\%R12C
		\bottomrule
	\end{tabu}		
\end{table}
The configuration \num{1} is our reference, corresponding to the complete model.
The optimized parameters are close to the ones from \autoref{tab:intrinsics}, \ie with less than \SI{1}{\percent} of variation, for all converging configurations and for all datasets.

First, the distortions do not impact the reprojection error of the model. 
Considering the pairs of configurations $\left(1,2\right)$, $\left(3,4\right)$, and $\left(5,6\right)$, the errors are similar with or without distortions, indicating that our camera does not suffer from lateral distortions.
This is due to the relatively large main lens focal length.
Nevertheless, distortions may have a role to play in case of shorter focal length.
Second, removing the rotations of the \ac{MLA} does not improve nor worsen the reprojection error and the pose estimation.
When keeping the tilt but freezing the pitch, the model is able to converge.
The tilt, in combination with other factors (such as a slight decrease of the main lens focal length), compensates for the error introduced by the approximate value of the pitch.
In contrast, configurations \num{7} and \num{8} do not converge to a solution, showing that when removing both the tilt and the pitch of the \ac{MLA}, the model is not constrained enough, and the reprojection error cannot be minimized, resulting in a failure. %

Finally, when freezing the pitch to its initial value, the positional part of the reprojection error increases. %
It is especially the case for dataset \texttt{R12-A}, where the reported errors in \autoref{tab:ablation} are the highest of all configurations.
This confirms our previous observation that the deviation of the micro-image centers and their optical centers does not satisfy an orthographic projection between the \ac{MIA} and the \ac{MLA}.
The pitch should be taken into account, on one hand to improve the precision of the model, and on the other hand not to hinder the optimization process.

\subsection{Relative blur calibration}

\begin{figure}[]
	\centering
	\includegraphics[width=\linewidth]{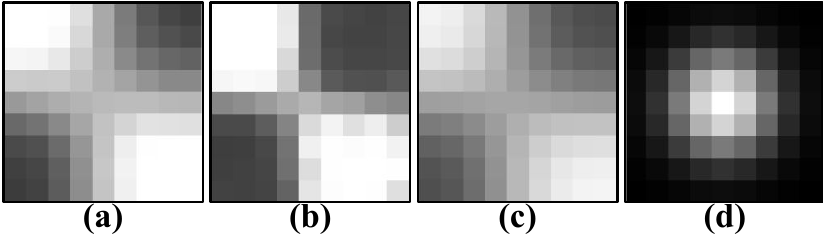}
	\caption{
		\textbf{(a)} Reference image with highest amount of blur.
		\textbf{(b)} Target image to be equally-defocused.
		\textbf{(c)} Target image with additional blur.
		\textbf{(d)} Estimated \acf{PSF}.
	}\label{fig:blur}\vspace*{-3mm}
\end{figure}

We calibrate the blur proportionality coefficient $\blurpropcoef$ for the three datasets using our \ac{BAP} features.
\autoref{fig:blur} presents two windows extracted around \ac{BAP} features of different types from the same cluster, showing different amount of blur. %
The target image to be equally-defocused according to our model is shown before, (b), and after, (c), blur addition.
The estimated \ac{PSF} of the relative blur is given in (d).
The optimized blur proportionality coefficients $\blurpropcoef$ are reported in \autoref{tab:intrinsics}.
Theoretically, the parameter should be the same for all three datasets.
Empirically this observation is validated for \texttt{R12-A} and \texttt{R12-B}.
Estimated $\blurpropcoef$ for \texttt{R12-C} is lower. %
This is because the micro-lenses focal lengths in \texttt{R12-C} are slightly shorter than in \texttt{R12-A} and \texttt{R12-B}.
Analytically, this difference generates a higher amount of relative blur, and thus a shorter estimate of $\blurpropcoef$ to match the observed blur in image space.
In other words, $\blurpropcoef$ compensates for the slight differences in $\focal\mltype{i}$ estimates. 
Therefore, $\blurpropcoef$ should be calibrated for each dataset.

\vspace*{-5mm}\subsection{Profiling the plenoptic camera}

\begin{figure}[t]
	\centering
	\includegraphics[width=\linewidth]{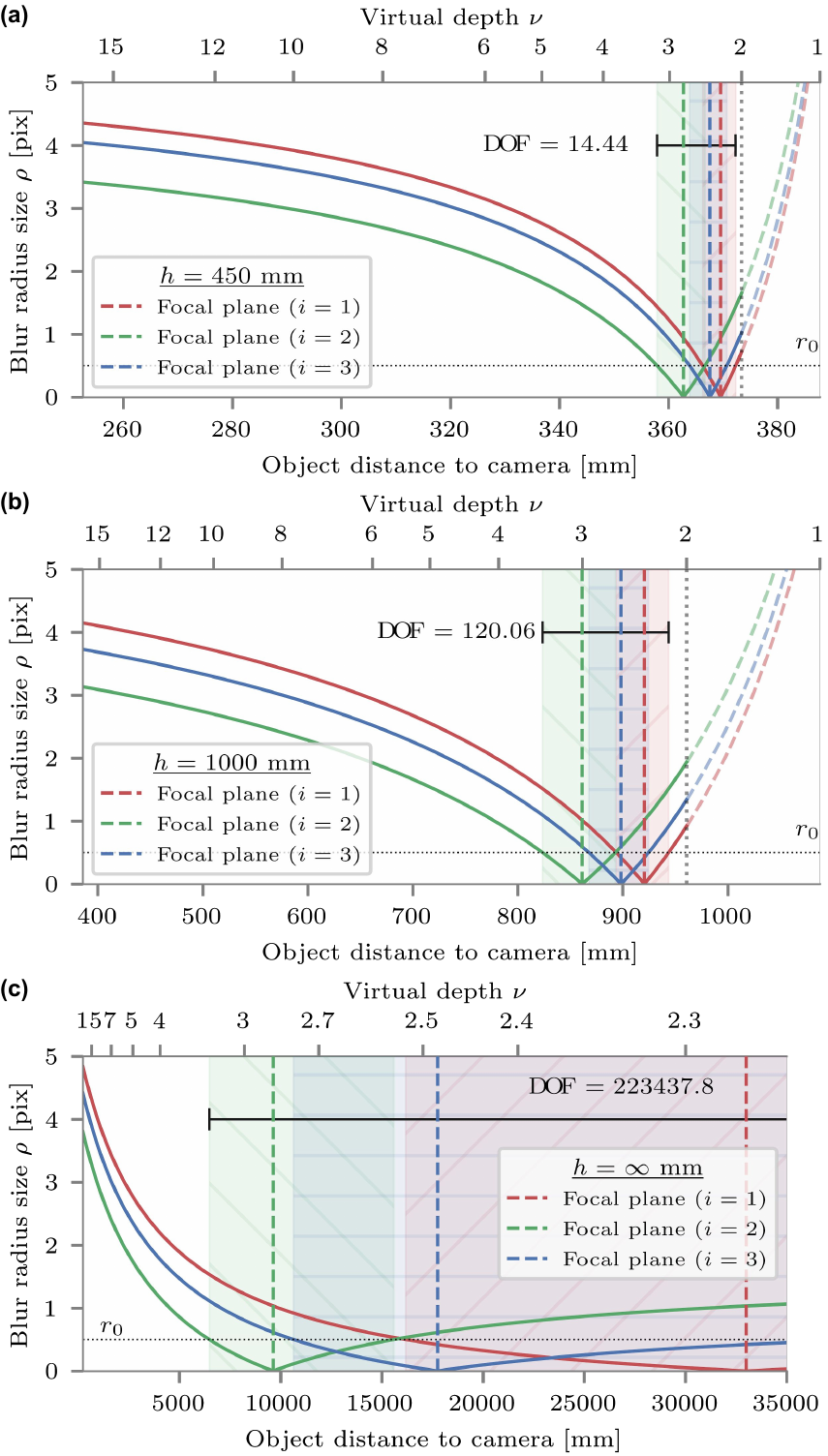}%
	\caption[Blur profiles of each micro-lens type]{%
		Blur profile, including each micro-lens type, in object space, at different focus distances: \textbf{(a)} $\focusdist = 450~\si{\mm}$; \textbf{(b)} $\focusdist = 1000~\si{\mm}$; and \textbf{(c)} $\focusdist = \infty~\si{\mm}$.
		Focal planes and \aclp{DoF} are illustrated for each type.	
		The blur radius is expressed in \si{pixel} as function of the object distance to the camera in \si{\mm}. 
		Corresponding virtual depth is reported on the secondary $x$-axis. 
	}
	\label{fig:blurprofileobj}\vspace*{-4mm}
\end{figure}

Using the parameters from our calibration process, we plot the \textit{blur profile} of the camera, \ie the evolution of the blur radius with respect to depth for each micro-lens type along with its corresponding \ac{DoF}. %
\autoref{fig:blurprofileobj} shows the blur profiles obtained for our three focus distance configurations, with their \acp{DoF} expressed in \si{\mm}. 
The blur radius is expressed in {pixel} and is given for each type, in \textit{red} for type $(1)$, in \textit{green} for type $(2)$ and in \textit{blue} for type $(3)$.
Distances are given in object space in \si{\mm} with their corresponding virtual depth on a secondary $x$-axis, spanning from $\virtdepth=1$ to $15$, except for the configuration $\focusdist = \infty$ where we cropped just after the farthest focal plane.
In \ac{MLA} space, the profiles have the same behavior for all focus distances, as it only depends on the \ac{MLA} parameters.

First, the horizontal dashed line represents the radius of the minimal acceptable \acl{CoC} $r_0$.
In our case, at a wavelength of \SI{750}{\nano\meter}, the radius of the smallest diffraction-limited spot is $r^* = 2.4~\si{\micro\meter}$ which is less than half the pixel size.
We then choose $r_0 = \pixelsize/2$.
Despite not illustrated in the figure, the blur radius grows exponentially when getting closer to the plane $\virtdepth = 0$. 
Once this limit is exceeded, the blur decreases and converges to a constant value of approximately \SI{6}{pixel}. 
This happens for more distant objects when points are projected in front of \ac{MLA} implying a negative virtual depth.
This is the case for $\focusdist=450$ and $\focusdist=1000~\si{\mm}$, but not for $\focusdist=\infty$, as the points were never projected closer than $\virtdepth=2$.
In the working distance range, the blur does not exceed \SI{5}{pixel} and grows when points are closer to the camera.
Secondly, we can use the \ac{DoF} to select the range of working distances where the blur is not noticeable.
The \ac{DoF} increases in object space as the focus distance increases. 
As reported on the figures: for \texttt{R12-A}, the \ac{DoF} is of $14.44~\si{\mm}$; for \texttt{R12-B} of $120~\si{\mm}$; and finally, for \texttt{R12-C}, the total \ac{DoF} is of $223~\si{\m}$.
In \ac{MLA} space the total \ac{DoF} is constant and spans from $\virtdepth = 2.15$ to $3.45$.
As expected, the \acp{DoF} overlap. 
In particular, the \ac{DoF} of the type $(3)$ micro-lens is entirely included in the other two, whereas the \acp{DoF} of the type $(1)$ and $(2)$ just touch.
Within the total \ac{DoF}, a point can then be seen focused in two micro-images of different types simultaneously, which eases the matching problem between views.

Finally, we can easily identify the distance limits at which the point will not be in the \ac{DoF} anymore nor be projected on multiple micro-images, \ie corresponding to virtual distances $\abs{\virtdepth} < 2$.
At these distances, disparity cannot be computed in image space, and no depth estimation can be performed.
Such estimation can also be hindered by the resolution in virtual space compared to the resolution in object space as disparity is inversely proportional to virtual depth.
For instance, for close objects, points will be projected on more micro-images but with a low disparity.
So the profiles can be used to efficiently characterize the range of distances according to the desired application.
Furthermore, once the \ac{MLA} parameters are available, we can simulate an approximate blur profile for the desired focus distance $\focusdist$ with the desired main lens focal length $\Focal$ by updating the value of $\Dist$ using \autoref{eq:initdd} and \autoref{eq:focusdist}.

\vspace*{-6mm}\section{Conclusion}\label{sec:conclusion}\vspace*{-2mm}

To calibrate a plenoptic camera, state-of-the-art methods rely on simplifying hypotheses, on reconstructed data or require separate calibration processes to take into account the multi-focus configuration.
Taking advantage of blur information we propose: 
1) a more complete plenoptic camera model with the introduction of a new \ac{BAP} feature that explicitly models the defocus blur; this new feature is exploited in our calibration process based on non-linear optimization of reprojection errors;
2) a new relative blur calibration to fill the gap between the physical and geometric blur, which enables us to fully exploit blur in image space;
and 3) a way to profile the plenoptic camera and its extended \acf{DoF}.

Our camera model is applicable to the multi-focus plenoptic camera (both in Galilean and Keplerian configuration), as well as to the single-focus and unfocused plenoptic camera.
In case of the \texttt{Raytrix} multi-focus camera, our ablation study shows that main lens distortions and \ac{MLA} tilt can be omitted without hindering the calibration process nor the pose estimation. 
The study also indicates that explicitly including the pitch of the micro-lenses in the model improves the results.
In addition, our calibration methods are validated by quantitative evaluations in controlled environment on real-world data.
Our method provides strong initial intrinsics during the pre-calibration step, and coherent optimized camera parameters for all evaluated configurations.
It shows a low and stable relative translation error across all the datasets.
In the future, we plan to use blur information in complement to disparity to improve metric depth estimation.

\vspace*{-4mm}\begin{acknowledgements}
We thank Charles-Antoine Noury and Adrien Coly for their insightful discussions and their help during the acquisitions.
\end{acknowledgements}

\section*{Declarations}
\section*{Data Availability Statement}

The datasets analyzed for this study can be found in the following repository: \url{https://github.com/comsee-research/plenoptic-datasets}.
All data supporting the results of this paper will be provided upon request.

\section*{Code availability Statement}\vspace*{-3mm}

The code source has been made open-source and publicly available in the following repositories: \url{https://github.com/comsee-research/libpleno}, and \url{https://github.com/comsee-research/compote}.

\section*{Conflict of Interest Statement}\vspace*{-3mm}

The authors declare that the research was conducted in the absence of any commercial or financial relationships that could be construed as a potential conflict of interest.

{\small
	\printbibliography
}

\end{document}